\documentstyle[prb,aps,epsf,epsfig,twocolumn]{revtex}
\def\gtappr{{{\lower4pt\hbox{$>$} } \atop \widetilde{ \ \ \ }}}
\def\ltappr{{{\lower4pt\hbox{$<$} } \atop \widetilde{ \ \ \ }}}
\def\be{\begin{equation}}
\def\ee{\end{equation}}
\def\bea{\begin{eqnarray}}
\def\eea{\end{eqnarray}}
\def\bk{{\bf k}}

\def\dg{{^{\dag}}}

\def\btau{\mbox{\boldmath $\tau$}}
\def\bone{\mbox{\boldmath $1$}}

\def\Tr{\text{Tr}}
\def\sgn{\text{sgn}\,}

\newcommand{\rarrow}{\rightarrow }

\def\etal{{\it et al.\/}}
\newlength{\wdy}
\newlength{\hty}
\newlength{\widf}
\newcommand{\figa}[3]
{\begin{figure}\begin{center}
\leavevmode\hbox{\epsfxsize=\widf \epsffile{{#1}}}
\end{center}
\caption{#2}\label{#3}\end{figure}}
\newcommand{\boxy}[3]
{\raisebox{-#2\columnwidth}{\epsfig{figure={#1},width=#3\columnwidth}}}

\begin{document}
\draft
\twocolumn[\hsize\textwidth\columnwidth\hsize\csname@twocolumnfalse%
\endcsname


\title{Magnetization of a D.C. biased quantum dot}
\author{P. Coleman and W. Mao}
\address{Center for Materials Theory, Rutgers University, Piscataway, NJ 08854, USA
}

\date{Dec 28, 2001}
\maketitle
\vskip 0.5 cm
\begin{abstract}

Using a quantum generalization of the 
Onsager principle of microscopic reversibility, 
the magnetization of a 
system in a non-equilibrium steady state quantum dot is formulated as
a response of the interaction energy  to an external field. 
This formulation permits a direct and compact computation of 
the steady-state magnetization of a non-equilibrium quantum dot
as a differential of  the interaction energy. Unlike the direct computation
of the magnetization using perturbative Keldysh methods, 
this approach does not
require the use of a point splitting procedure.
Our results nevertheless 
support earlier  calculations made in the limit of zero field,
and they support the 
survival of strong coupling
to arbitrarily large voltages, both at zero field, and under the conditions
where the chemical potential difference 
$eV$ becomes equal to the spin-flip energy in a field $eV = g \mu_{B }B$.

\end{abstract}

\vspace{1 cm}

\vskip -0.4 truein
\pacs{PACS numbers: }
\vspace{0.5 cm}
]
\narrowtext

\section{Introduction}

Quantum dots\cite{Intro} offer a unique opportunity to study the
behavior of a simple, yet non-trivial 
strongly correlated electron system out of equilibrium.\cite{Ng:1988,Glazman:1988,Meir:1992,Meir:1993,Wingreen:1994,GG:1998,Cronenwett:1998,vanderWiel:2000,Glazman:2000,Langreth,Schiller:1998,Avishai:1998}
In the Coulomb blockade regime, quantum dots with odd numbers of
electrons 
develop a magnetic moment that interacts
non-perturbatively with the leads at low temperatures. 
The unbiased quantum dot provides
a classic example of strongly correlated electron physics: 
at low temperatures,the
residual interaction of the dot spin with electrons in the leads gives
rise to a Kondo effect, whereby the formation of a many -body
resonance at the Fermi energy drives the conductance in the Coulomb
blockade regime back up to the unitary limit. 
The response of the Kondo effect to a D.C. bias is a matter of some
interest, and 
provides us with a chance  to  examine
how a non-trivial strongly correlated electron system responds in a
non-equilibrium setting. 

One of the key properties of the Kondo effect in equilibrium is the
presence of a ``running
coupling constant''\cite{poor} 
whereby the antiferromagnetic
coupling between the dot spin and the leads $g (\Lambda )$ grows
progressively as the energy scale $\Lambda $ is reduced. The Kondo effect
exhibits the phenomenon of confinement. At high energies,
the local moment is asymptotically free, weakly interacting with the
surrounding environment, but 
at low energies the spin of the dot is ``confined'':
screened by the lead electron.
A single scale, the ``Kondo temperature'' $T_{K}$, governs the low temperature
properties;
$T_{K} = D \sqrt{g} \, e^{-1/2g}$,
where $g$ is the `bare coupling' 
between the 
spin of the dot and 
the leads and $D \, {\gg} \, T_{K}$ is the electron bandwidth.
Thus, for example, the
magnetization at temperature $T$ and magnetic field $B$ is
a universal function
$M = m (T/T_K,B/T_K)$,
where $M$
has a perturbative ``weak coupling'' expansion 
in $g$ only when $T, B \gg T_K$. 
(See Fig.~1(a).) 

What happens to the Kondo effect when a current flows through a
D.C. biased
quantum dot? 
Some
have suggested 
that the passage of electrons through the
quantum dot will ``decohere'' the physics  of the Kondo effect\cite{Wingreen:1994,Kaminski:1999,roschprl,newschiller}, 
dephasing 
the coherent spin-flip processes necessary for the Kondo screening 
process.  The rate at which electrons pass through the quantum dot
is given by $I/e$, where $I$ is the current through the dot. 
The formation of a Kondo resonance requires 
that quantum processes are coherent on a time scale $\sim \hbar
/T_{K}$. 
Suppose that quantum 
processes on time-scales in excess of $\tau_{c} \sim e/I$
are dephased by the passage of electrons, then when $\tau _{c}\ltappr
\hbar /T_{K}$, or 
when
\[
I \gtappr \frac{e T_{K}}{h}
\]
the coherent formation of a Kondo resonance is expected to break-down.
This would imply that the Kondo effect will break down
at voltages in excess of the Kondo temperature. 
In such a picture, the magnetization of the quantum dot
\[
M= M (T/T_{K}, H/T_{K}, eV/T_{K})
\]
should become perturbative 
in the coupling to the leads at arbitrarily low temperatures once 
$eV>>T_{K}$. 

An alternative picture of the quantum dot  argues 
that the Kondo effect is modified, but not dephased by the passage of electrons
through the dot\cite{Wen:1998,us}. 
According to this picture, 
although interlead Kondo processes are cut-off by a 
finite voltage, intralead Kondo renormalization effects continue unabated 
in each lead until 
the system enters a non-equilibrium strong-coupling regime  at a
renormalized temperature $T_{K}^{*} (V)$. 
If true, the DC biased quantum dot should display a Curie
like susceptibility characteristic of a decoupled local moment even in
the presence of a curent, 
until the temperature approaches the renormalized Kondo temperature
$T_{K}^{* } (V)$. 
\widf=0.5\columnwidth\\
\figa{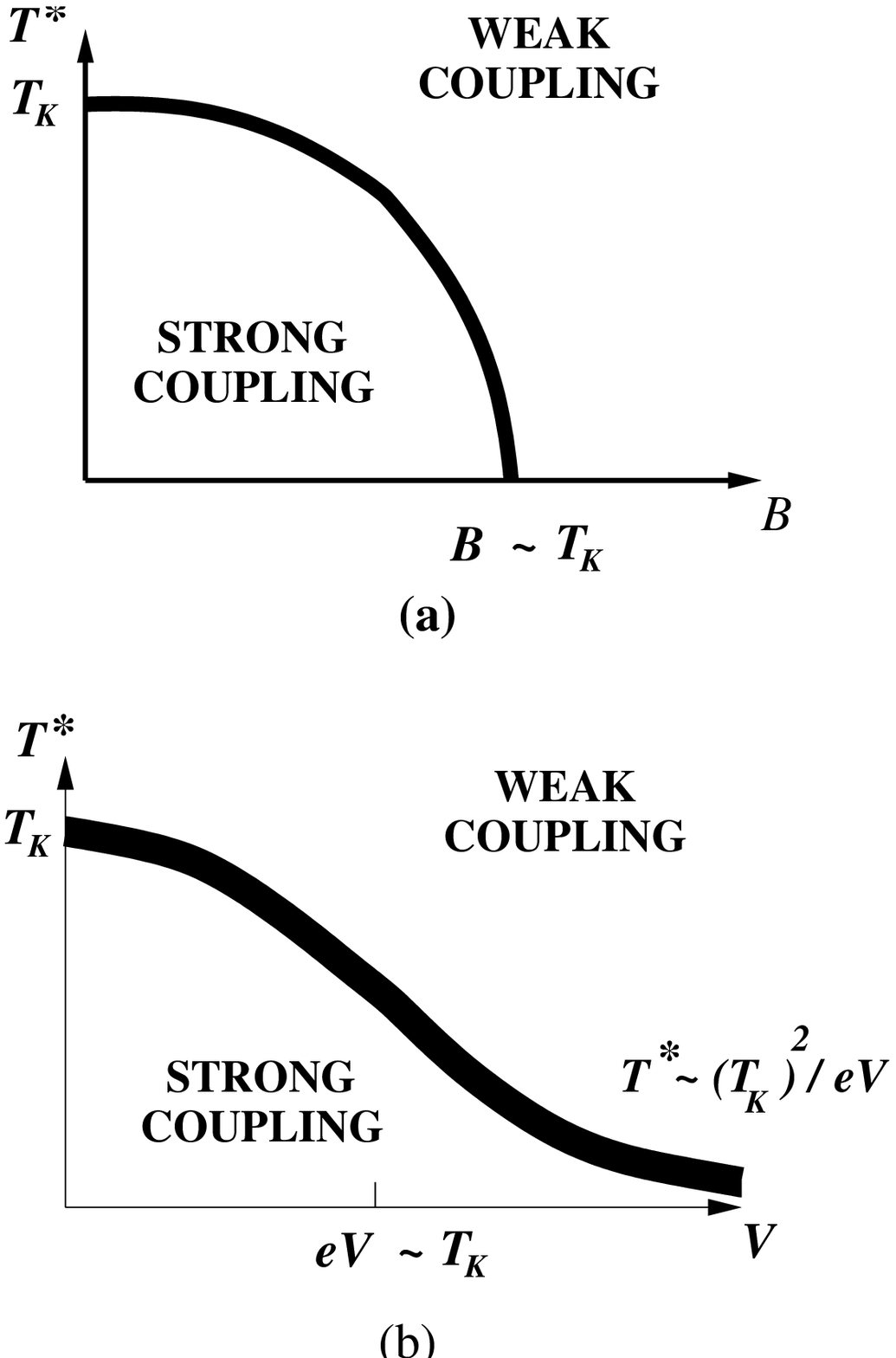}{(a) Field dependence of the crossover temperature $T_K^*(B)$ separating 
weak- and strong-coupling regimes of the equilibrium Kondo model.
At absolute zero, for fields larger than the Kondo temperature
the quantum dot re-enters weak coupling. 
(b) Voltage dependence of the 
crossover temperature $T_K^*(V)$, from earlier results of Coleman,
Hooley and Parcollet\cite{us}.  Note that $T_K^*$ goes
to zero only in the limit $V \to \infty$, i.e.\ for low enough temperatures, the model
reaches a strong coupling state irrespective of the magnitude of the
voltage.}
{genpic}\noindent 
Such a possibility rests 
on the observation that 
the Kondo screening process partially delocalizes the spin of the
quantum dot into the leads to form a kind of ``meso-spin'' \[
\vec{S}^{*} (\Lambda)= U (\Lambda )\vec{S}U\dg (\Lambda )
,
\]
whose composite structure depends on the scale
on which it is observed. Here, the idea is that same unitary
transformation $U (\Lambda )$ which integrates out the high energy electrons 
also admixes and hence delocalizes the bare dot spin
$\vec{S}$ into the leads. 
(Fig. 2.)
In this picture, the rapid fluctuations 
of the bare 
quantum dot spin $\vec{S}$ at a rate comparable with $\tau^{-1} _{c}$,
are merely internal redistributions of spin within the scale dependent
meso-spin, 
\[
S_{\uparrow}  \rightleftharpoons \left\{ \hbox{high energy singlet}
\right\}+ \{e,h \}_{\uparrow}.
\]
\vskip -0.2truein
\begin{figure}
\begin{center}
\leavevmode
\hbox{\epsfxsize=6cm \epsffile{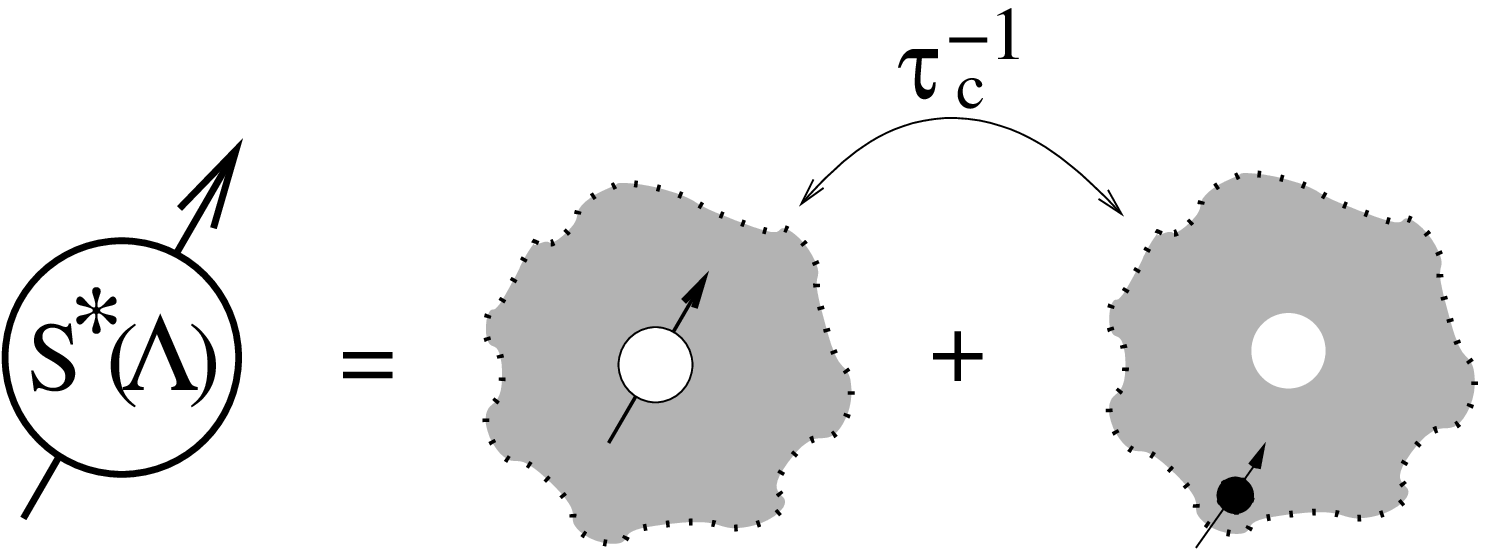}}
\end{center}
\caption{
Renormalized ``meso-spin'' which forms around
a magnetic ion as high-energy electrons are systematically integrated
out of the spectrum. The original spin of the bare local moment is
partially delocalized into the conduction sea. 
Fluctuations of the bare moment at a rate $\tau_{c}
^{-1}$
lead to an
internal
redistribution of spin within the meso-spin, but do not cause 
the composite object to flip. 
}\end{figure}
\noindent 
If this picture is correct, 
the magnetization of the DC biased quantum dot is expected to remain 
perturbatively close to a Curie law  at large voltages down to a 
renormalized Kondo temperature $T_{K}^{*}(V)$, where a new kind of
non-equilibrium Kondo effect will take place. 

These discussions suggest that a careful computation of the 
quantum dot magnetization may provide a key to understanding whether
a Kondo effect takes place in the quantum dot at a large voltage bias.
In this paper, we present an extended discussion of earlier
perturbative calculations of the quantum dot 
magnetization by Coleman, Hooley and Parcollet\cite{us} that suggest that
strong coupling physics extends to arbitrarily high voltages in the
D.C. biased Kondo model (Fig.~1 (b)).  Two of the three original
authors of that work, Parcollet and Hooley \cite{PH}
have recently argued that the key results are incorrect, 
claiming that 
the perturbative expansion of the magnetization breaks down at
zeroth order in the coupling constants 
when a current flows through the quantum dot.  
Part of the reason for doubts about
the earlier work stemmed from the necessity of introducing a
``point-splitting'' procedure to control divergences in the
perturbation theory associated with the conservation of the total
magnetization.  In this paper we present new arguments in favor 
of the original conclusions of Coleman, Hooley and Parcollet \cite{us}. 
Our arguments avoid some of the earlier difficulties associated with
point-splitting by taking advantage of a quantum mechanical extension
of Onsager's
principle of microscopic reversibility\cite{Onsager:1931a,Onsager:1931b}
to steady-state
non-equilibrium quantum mechanics. 

\section{Model and key results}
Under conditions where the quantum dot has almost integral valence,
interactions between the local moment on the dot and its leads
can be described by a Kondo model, of the following form:
\begin{eqnarray}\label{hammy}
H &=& \sum_{m {\bf k}\sigma} \varepsilon_{{\bf k}} c^\dag_{
m{\bf k} \sigma} c_{ m{\bf k} \sigma} + {\mathcal H}_{I}- \ B
\ M_{TOT},\cr
{\mathcal H}_{I}&=& J\psi \dg _{\alpha }
(0)\vec{\sigma }_{\alpha \beta }\psi _{\beta }\cdot {\vec{S}}.
\end{eqnarray}
Here, $c^\dag_{\alpha {\bf k} \sigma}$ creates an electron in lead
$m \in \left\{L,R\right\}$ with momentum ${\bf k}$ and spin
$\sigma$, and $J$ is a positive (antiferromagnetic) 
Kondo coupling constant derived from
virtual charge fluctuations on and off the quantum dot. 
For simplicity the spectrum of the leads  on either side of the
dot
is taken to be identical and 
\begin{equation}\label{}
\psi _{\alpha } (0)= \sum_{\vec{k}}\alpha _{L}c_{L{\bf k}\alpha  }+
\alpha _{R}c_{R{\bf k}\alpha  }
\end{equation}
describes a linear combination of electrons in the left and right
hand-leads
with $\alpha _{R}^{2}+\alpha _{L}^{2}=1$.
In equilibrium the left and right-hand leads are filled to the same chemical
potential $\mu$, but out-of-equilibrium, the left and right hand leads
differ
by an amount equal to the driving voltage
\begin{equation}\label{}
\mu_{L}-\mu_{R}= e V.
\end{equation}
Finally, note 
that we have included the magnetic field term in the interaction part of the
Hamiltonian,
where
\[
\hat M_{TOT} = 2 S_{z}+
\hat  M_{leads}
\]
is the total magnetization of the leads plus dot, 
\[
\hat M_{leads}=\sum_{{\vec{k}},\lambda \sigma } \biggl[
(n_{\lambda {\vec{k} }\uparrow}-n_{\lambda {\vec{k} }\downarrow})
\biggr] 
\]
is the magnetization of the leads and 
$n_{\lambda {\bf k} \sigma }= c\dg _{\lambda {\bf
k}\sigma }
c _{\lambda {\bf k}\sigma }
$ are  the number operators for electrons in the leads.

The interaction term in this Hamiltonian can divided up into inter and
intra-lead terms, as follows
\begin{equation}\label{ham}
{\mathcal H}_{I} =H_{\rm refl} + H_{\rm trans},
\end{equation}
where
\begin{eqnarray*}
H_{\rm refl} & = & J_{R} 
\sum_{{\bf k},{\bf k}',\sigma,\sigma'}
\left(
c^\dag_{R{\bf k}\sigma} {\vec \sigma}_{\sigma \sigma'}
 c_{R {\bf k}' \sigma'} \right)
\cdot {\vec S} \,\, + \,\, (R \to L)  \\
H_{\rm trans} & = & 
J_{LR} 
\sum_{{\bf k},{\bf k}' , \sigma,\sigma'}
 \left(
c^\dag_{R{\bf k}\sigma} {\vec \sigma}_{\sigma \sigma'} c_{L {\bf k}' \sigma'} \right)
\cdot {\vec S} \,\, + \,\, (R \leftrightarrow L) 
\end{eqnarray*}
describe the spin-flip reflection and transmission of electrons
between
the leads and the coupling constants
\begin{eqnarray}\label{cupc}
J_{L}&=& \alpha_{L}^{2}J, \qquad 
J_{R}= \alpha_{R}^{2}J\cr
J_{RL}&=&J_{{RL}}=\alpha _{R}\alpha _{L}J
\end{eqnarray}
It is sometimes useful to consider the coupling constants
$J_{L}$, $J_{R}$ and $J_{RL}$ as independent variables. 

In equilibrium, the Kondo model has been solved exhaustively by
a number of different methods.  Central to our whole understanding
of the Kondo model, is the renormalization group. Unfortunately, a 
complete framework for the renormalization group out-of-equilibrium
is not yet available.   Much of the current understanding of the
Kondo model out-of-equilibrium is based on resummed perturbation
theory. A second approach to the problem has been to use
a strong-coupling analysis.  Each of these approaches is faced
with its own particular bias. In a resummed perturbation theory,
certain classes of processes are selected, and there is genuine
danger that cancellations inherent to the full problem are lost
in the process of selecting diagrams. By contrast, 
the strong coupling approach relies heavily on the linearization of
the band, and the assumption that the pertinent band-electron cut-offs
in strong coupling are larger than the applied voltage.  
However, we know from the equilibrium  Kondo model, that in some sense,
the band-width of the strong-coupling problem is given by the Kondo
temperature, so it is not clear how this last assumption can be
reconciled with  a voltage larger than the Kondo temperature.

In the face of these uncertainties, we argue that the only unbiased
way to address the issue of whether the DC biased Kondo model
enters a strong-coupling regime, is to revisit perturbation theory in 
a non-equilibrium context.  There are two obvious physical variables that
can be used to characterize the Kondo effect out-of-equilibrium: the
current through the dot, and the total magnetization.  Perturbative
studies of the current\cite{Glazman:2000} through a quantum dot indicate that the
expansion
\[
I = \frac{e}{\hbar }f (T/T_{K},eV/T_{K})
\]
enters a weak coupling, perturbative regime when $eV>>T_{K}$. At first
sight, this might be taken as evidence that the Kondo effect returns
to weak coupling at large voltage bias.  However, the current between
two leads at large voltage bias $eV>> T_{K}$ involves electrons far
from the Fermi surface of each lead, and these electrons would be
weakly
coupled to the dot spin even under equilibrium conditions.

We shall argue that a 
more useful quantity to characterize the behavior of a quantum 
dot out of equilibrium, is the static magnetization.  
We shall take the magnetization $M$
to refer to the impurity contribution to the {\sl total }
magnetization, defined as the difference 
\[
\hat  M= \langle \hat  M_{TOT}\rangle_{J} - \langle M_{leads}\rangle_{J=0} .
\]

In the region of weak coupling $max (T,H)/T_{K}>>1$, the equilibrium 
magnetization
has a perturbative expansion. For low fields,  $M (T)= \chi (T)B$, where
\begin{eqnarray}\label{}
\chi (T) = \frac{1}{T}\left[1 - 2J\rho - 4 (J\rho )^{2}\left(
\log (\frac{De^{(3/4 +\gamma )}}{2 \pi T}) \right) \dots  \right]
\end{eqnarray}
where $\gamma =0.5772=-\psi (1)$ is Euler's constant, $D$ is the band-width and
$\rho $ is the density of states in each lead.
In the same cut-off scheme, the high-field magnetization $B>>T$ is given by
\begin{equation}\label{hunk}
M = 1\left[1 - J\rho - 2 (J\rho )^{2}\ln (\frac{D}{2B})- \dots \right]
\end{equation}
The cross-over to strong-coupling is defined by the scale at which
the second and first order corrections to the magnetic susceptibility
and magnetization become equal. This defines a field scale
\begin{equation}\label{A}
2B\equiv  T_{K}=D\exp [-1/ (2 J \rho) ] 
\end{equation}
and a temperature scale, 
\begin{equation}\label{B}
T_{o}= D\frac{e^{\frac{3}{4}+\gamma }}{2 \pi }\exp [-1/ (2 J \rho) ] =
0.60 T_{K}\end{equation}

In this paper, we generalize these results to the case of finite DC
bias, computing the magnetization to second order in the coupling constant
at finite temperature, field and voltage. In order to do this, we need
to define what we mean by the magnetization.  We consider the effect
of turning on the the Kondo coupling constant and computing how the
magnetization changes in the steady state, as we slowly increase the
coupling constant up to its maximum value. We shall argue that under rather
general conditions, the change in the magnetization due to the Kondo
interaction can be written in the form
\[
\Delta M = - \frac{\partial}{\partial B}
\int_{0}^{J} \frac{dJ'}{J'} \langle {\mathcal H}_{I}\rangle _{J',H}
\]
where $\langle {\mathcal H}_{I}\rangle _{J,H}$ is the steady state value of the
interaction energy, evaluated at Kondo coupling constant $J$, in a
magnetic field $B$. The advantage of this formulation, is that
it permits a ready generalization of equilibrium methods to a non-equilibrium,
steady state situation. 

\widf= 0.6\columnwidth
\figa{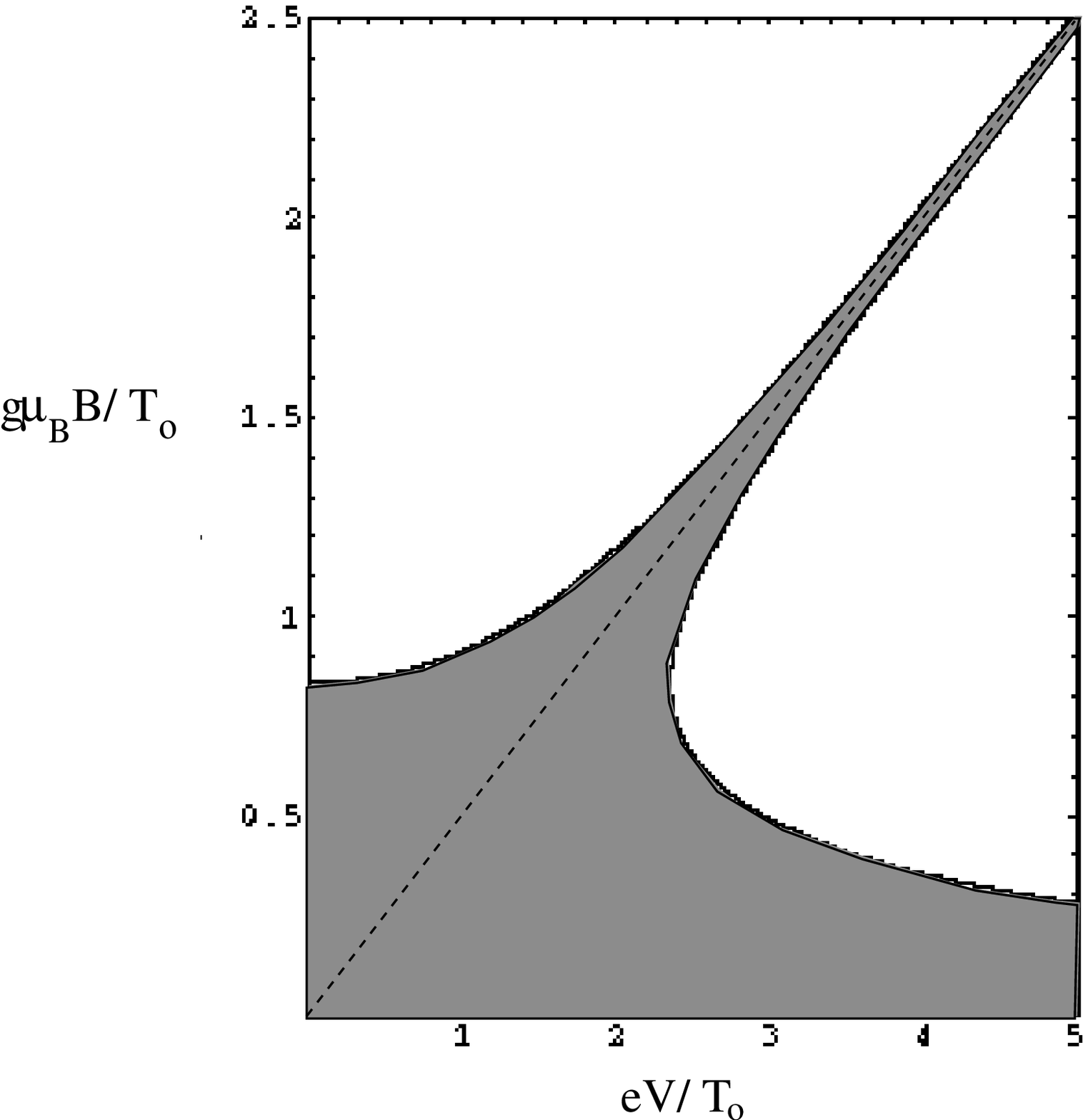}{Showing the region of strong coupling defined
by the equality of the first and second order terms in the expansion of
the magnetization.  Strong coupling persists to arbitrarily high voltage
at $B=0$ and for $g\mu_{B}B\sim eV$.}{strcoupl}
\vskip 0.3truein
The main result of our calculation is succinctly captured by the 
zero temperature magnetization, which for 
the symmetric case 
$J_{R}=J_{L}=J_{RL}= J/2$, takes the form
\begin{equation}\label{hunk2}
M = \left[1 - J\rho -  (J\rho )^{2} \ln \left(
\frac{D^{2}}{2B
\sqrt{\vert (eV)^{2}-(2B)^{2}
\vert}
}
\right)  \dots \right]
\end{equation}
This expression is perturbative so long as 
\begin{equation}\label{scoup}
 2B \sqrt{\vert (eV)^{2}- (2B)^{2}\vert }\gtappr
( T_{K})^{2}
\end{equation}
which defines  the region shown in Fig. 2.

As can be seen from Fig. 2., large voltage identifies two regions of strong-coupling:
\begin{itemize}
\item  Zero field Kondo effect  $2B < ( T_{K})^{2}/eV $,
corresponding to a  strong 
coupling region associated with the intra-lead Kondo processes that
are not cut-off by a finite voltage.  Earlier work\cite{Wen:1998,us}
suggested the corresponding fixed point is a two-channel Kondo model
where 
\[
H_{I}^{*}
\sim 
\left[
J_{R}^{*}
\psi \dg _{R}\vec{\sigma }\psi
_{L}+
J_{L}^{*}
\psi \dg _{L}\vec{\sigma }\psi
_{L}
\right]\cdot S
\]
More recent work by Rosch et al \cite{roschprl} and
Schiller et al \cite{newschiller} has suggested that
the entry into the strong-coupling 
two-channel regime is ultimately cut-off by
decoherence effects.  We shall return to this point in the discussion.

\item Finite field  Kondo effect. This occurs around $eV\sim 2B \equiv  g\mu _{B}B$,
$\vert eV -2B\vert< T_{K}^{4}/ (eV)^{3} $, and also extends up to
arbitrarily large voltage. Here the Kondo effect 
involves the degeneracy between a spin flip of the quantum dot 
and the transfer of an
electron between the fermi surface of the two leads.  
The effective Kondo
model for this situation presumeably resembles a one-channel 
Kondo model at {\sl zero voltage bias}, 
where the down electrons on 
high voltage lead and the ``up'' electrons on the low voltage 
lead  can scatter elastically between leads by exchanging spin with
the
local moment.  We expect that the physics at energy scales lower than $\Lambda
\sim \hbox{max} ( g\mu_{B }B,eV) $, will involve only these 
electrons, and the effective model will be 
a one channel Kondo model of the form 
\begin{equation}\label{}
H_{I}^{*}\sim J^{*}\vec S \cdot 
(\psi \dg _{R\uparrow},\psi \dg _{L\downarrow })\vec{\sigma }
\pmatrix{\psi _{R\uparrow}\cr \psi _{L\downarrow }
}
\end{equation}
where the spin $S$ sits in a zero field, and the
chemical potential difference  between the $L\downarrow $ and
$R\uparrow$ electrons is $\Delta \mu ^{*}= \Delta \mu -g\mu_{B}B $.  
In this situation, we
expect the spin on the quantum dot to be
perfectly quenched at low temperatures . Small changes in the voltage
will affect magnetization, 
however, without a mechanism for the transfer of electrons from one lead
to another, this finite field Kondo effect will not, it appears, lead
to a unitary differential conductance. 
\end{itemize} 

At finite temperatures, our general result takes the form 
\begin{eqnarray}\label{}
M &=& (1- \bar  J)\tanh \left(B \bigl[1-\bar J\bigr]/T \right)-
\cr
&& 2 [(\bar J_{R})^{2}+(\bar J_{L})^{2}]
 \frac{\partial }{\partial B}
\left[M_{o}B \left(\ln \frac{D}{2 \pi T} - \phi (\frac{2B}{T},0)\right) \right]
\cr
&-& 4(\bar J_{RL})^{2}
 \frac{\partial }{\partial B}
\left[M_{o}B \left(\ln \frac{D}{2 \pi T} - \phi (\frac{2B}{T},\frac{V}{T})\right) \right]+O (\bar J^{3})
\end{eqnarray}
where $\bar J_{\alpha }= J_{\alpha }\rho $, $\bar J = (\bar J_{R}+
\bar J_{L})$ and
\begin{eqnarray*}
\phi (b,v)&=& Re \int dx \frac{1}{(2 \cosh (\frac{x}{2}))^{2}}\times \cr
&&\frac{1}{4}\sum_{\sigma ,\gamma = \pm 1}  
2 \pi i \sigma \ln \tilde{\Gamma} (x+\sigma b+\gamma v)
\cr
\tilde \Gamma (x)&=& \Gamma (\frac{1}{2}+\frac{x}{2
\pi i})]
\end{eqnarray*}
\vskip 0.1truein
\begin{figure}
\begin{center}
\leavevmode
\hbox{\epsfxsize=6cm \epsffile{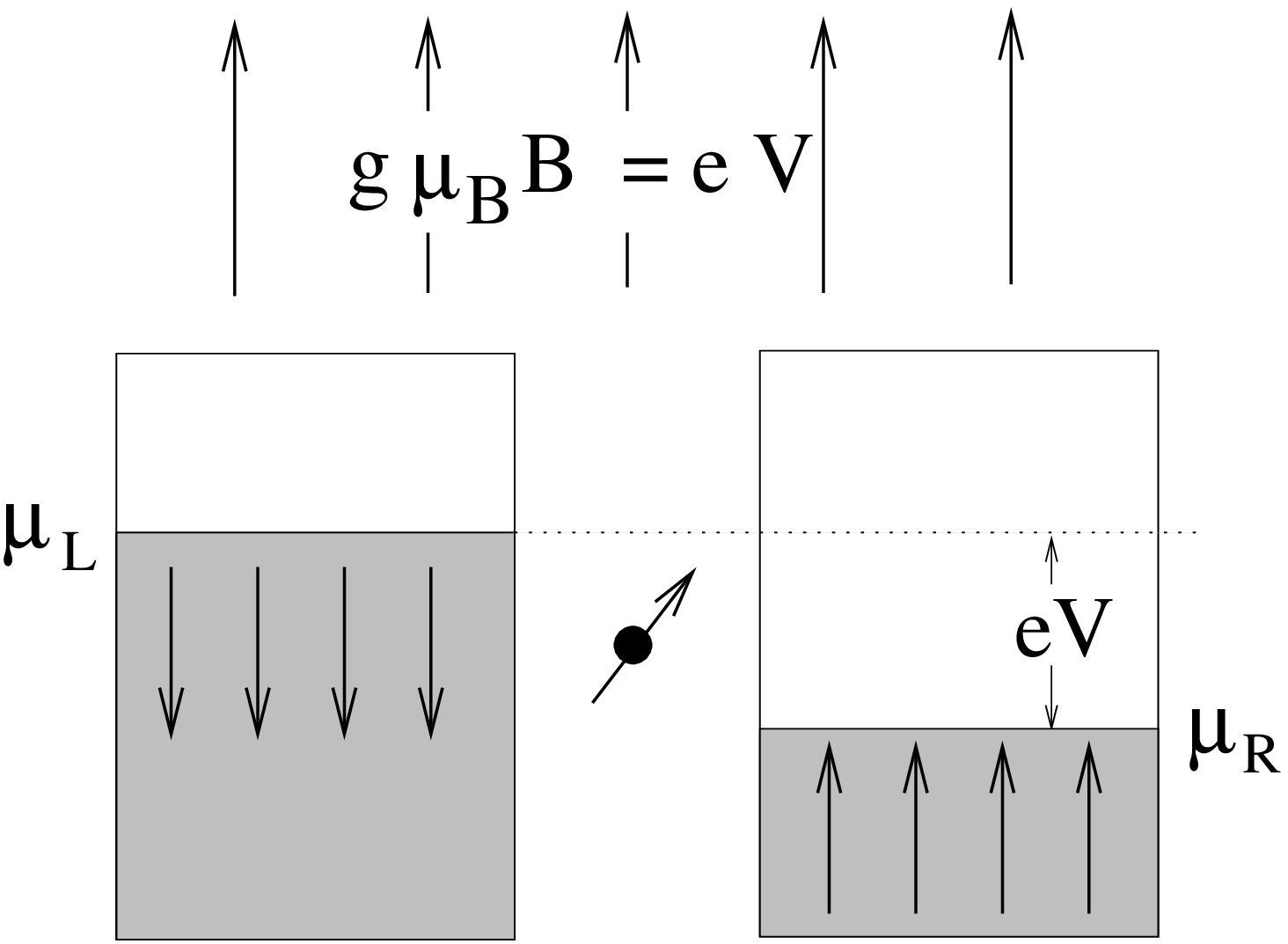}}
\end{center}
\caption{
Illustrating the Kondo effect
at a finite field $g\mu_{B }B = e V_{o}$.  At absolute zero,
we expect the ground-state to be a singlet and small departures in the
voltage from $V_{0}$ will then lead to a magnetic polarization 
of the local moment. 
}\end{figure}

In the zero field, finite temperature limit, this general result
reverts to the form $M=\chi B$, where
\begin{eqnarray}\label{suscept}
\chi_{i} &=& \frac{1}{T}
\left\{
 1 - 2(\bar J_{R} +\bar J_{L}) -
4\left(
   \bar J_L^2 + \bar J_R^2
\right)
\ln \left(
       \frac{De^{\frac{3}{4}+\gamma}}{2\pi T 
}
    \right) 
\right. 
\cr
&&\left.
- 8|\bar J_{LR}|^2 \left[ \ln
 \left(
\frac{De^{\frac{3}{4}+\gamma}}{ 2\pi T 
} 
\right)
- \Phi  \left( \frac{V}{T} \right)
\right] 
\right\}.
\label{pertu}
\end{eqnarray}
where $\gamma=0.5772\dots = - \psi (1)$ is the Euler constant.  The
crossover function $\phi(x)$, in terms of digamma
functions $\psi(z)$, is
\begin{equation}\label{DefPhi}
\Phi (x)  =
\hbox{Re} \int_{-\infty }^\infty 
\frac{dy}{4 \cosh^2 \bigl(\frac{y}{2}\bigr)}\left[
\tilde{\psi } (y+x)-\tilde{\psi}(y) \right],
\end{equation}
where $\tilde{\psi}(x)=\psi ( \frac{1}{2} + i \frac{x}{2\pi} )$.

The second-order terms in (\ref{pertu}) describe the leading
logarithmic enhancement of the Kondo coupling. 
Terms of order $J_{LR}^2$ involve inter-lead
processes and, as expected, the 
logarithmic divergence in these terms 
is cut by the voltage.  (To
see this, note that $-\phi (V/T)\sim -\ln (V/T)$ for $V \gg T$ which 
cancels the logarithmic temperature divergence.) 
By contrast, the intra-lead terms of order $J_R^2$ and $J_L^2$
are completely unaffected by the voltage $V$, which guarantees that the
leading logarithmic divergence survives
at {\em arbitrarily high voltage}.  This is easily seen in the large-$V$
form of the susceptibility,
\begin{eqnarray}\label{largeVsuscept}
\chi_{i} &=& \frac{1}{T}
\left\{
 1 - 2(\bar J_{R} +\bar J_{L})  -
4 \left ( 
       \bar J_L^2 + \bar J_R^2
 \right)
\ln \left(
        \frac{De^{1+\gamma}}{2\pi T 
}
    \right)
\right.
\\
&&\left.
- 8|\bar J_{LR}|^2  \ln \left( \frac{D}{V} \right)
\right\}, \qquad \quad V \gg T.
\end{eqnarray}
The survival of the leading logarithms in the susceptibility at
arbitrarily high voltage is consistent with the 
zero-temperature, finite field results, and at least at this order in
perturbation theory, is a signature that the {\sl intra}-lead Kondo effect 
continues unabated at temperatures or fields smaller than the voltage. 

The corresponding
boundary where
weak-coupling perturbation theory breaks down is shown in Fig. 4. 
\begin{center}
\epsfig{figure=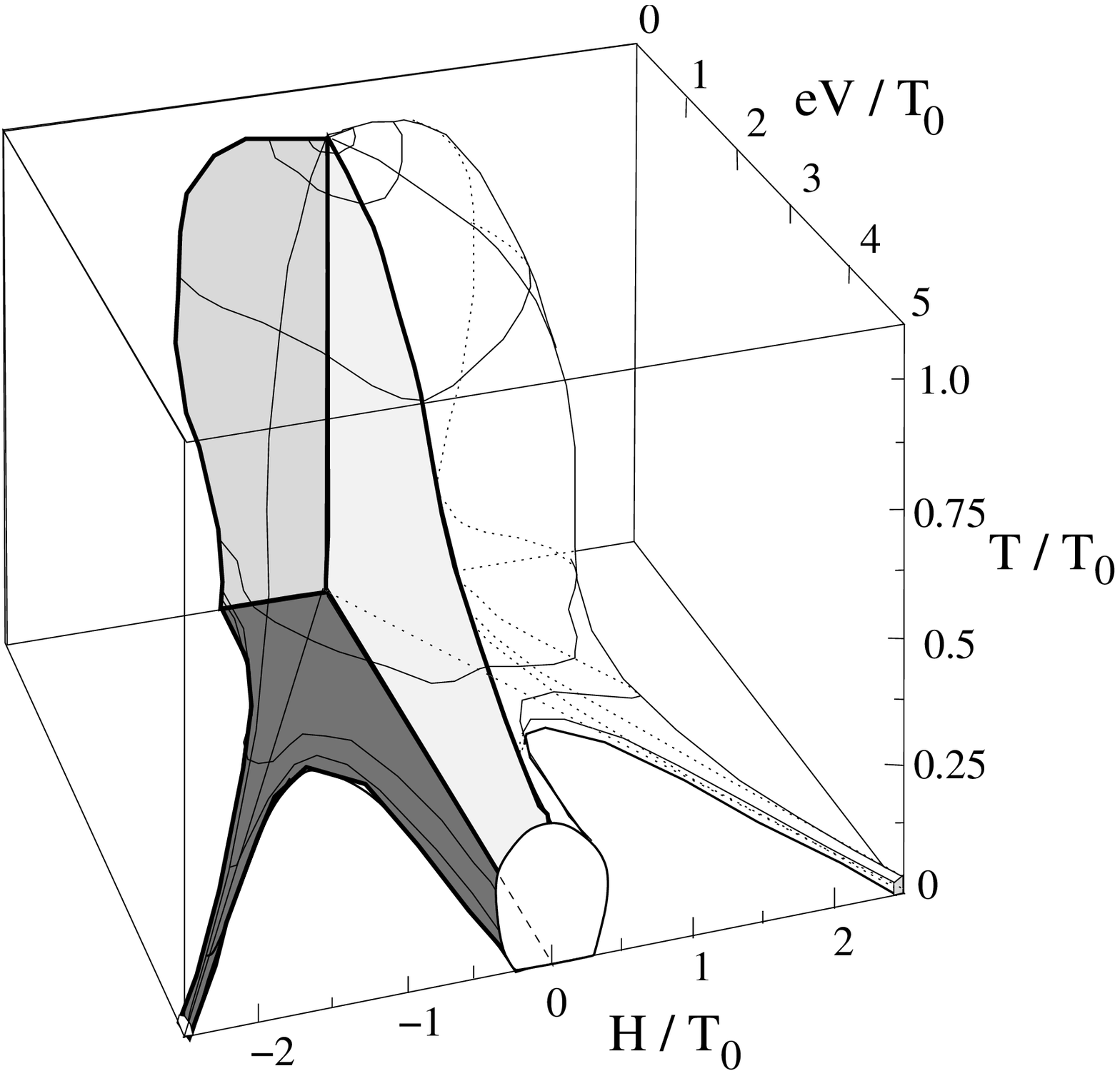,width=0.75 \columnwidth}
\end{center}

\vskip 0.1truein
{\bf Fig. 4. }Illustrating region of strong coupling for finite field, temperature and
voltage. 

\section{Magnetization as  a Response Function}

Since we wish to consider the case where the voltage $V_{\rm sd}$ takes an arbitrarily
large value, linear response methods are not appropriate; instead, we must resort to
the Schwinger-Keldysh formalism for non-equilibrium field theory.
The SK method\cite{Rammer} divides the Hamiltonian of the system into two parts:
\be
H = H_0 + H_{I}+ \lambda (t) A,
\ee
where, for convenience we have introduced a source term with field
$\lambda (t)$ coupled to quantity $A$.
It is assumed that we know the exact Green's functions for the
operators under $H_0$. The interaction $H_{I}$ and source term
are regarded as perturbations. 

\begin{figure}
\begin{center}
\leavevmode
\hbox{\epsfxsize=8cm \epsffile{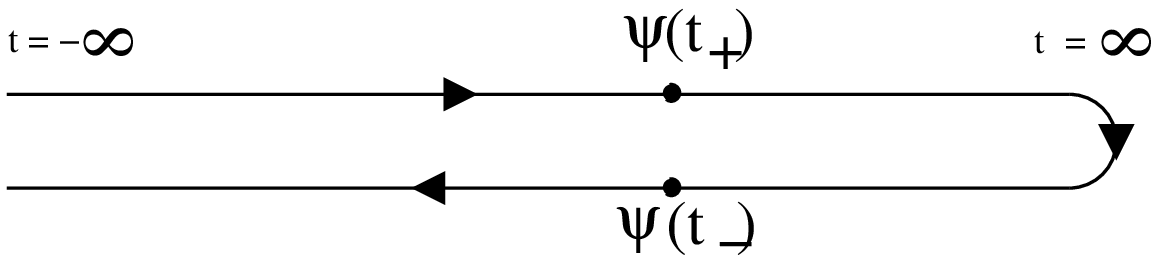}}
\end{center}
\caption{
The Keldysh contour.  The closed contour eliminates the need to
normalize the path-ordered exponential.  Fields are defined
on both the upper ($\psi(t_{+})$) and  lower ($\psi (t_{-})$) contour.
Along the path, the time $t_{-}$ is taken to be ``later'' than the
time
$t_{+}$.
}
\label{contours}
\end{figure}

In the SK approach, the expectation value
of an operator  $A (t_{o})$  in the interaction representation 
$A (t_{o})= e^{i H_{o}t_{o} }
Ae^{-i H_{o}t_{o} }
$ is 
\be
\langle A (t_{o}) \rangle = \langle P \left(A (t_{o}) e^{- i
\int_{K}dt \biggl[H_{I}+ \lambda (t) A (t)\biggr]}  \right)
\rangle 
\label{average}
\ee
where $K$ denotes the two branch Keldysh contour running from $-\infty
$ to $+\infty $ back to $t= -\infty $ ,
(Fig.~\ref{contours}b) $P$ denotes path ordering along this path and
 $\langle \dots \rangle = \Tr [\rho _{o}\dots ]$ is 
average with respect to the density matrix $\rho _{o}$
( in the absence  of the interaction) in the distant past.

In the conventional SK approach, the expectation value of
some quantity is computed directly, using diagrammatic methods.
We may regard this as a direct ``measurement''of the quantity in question.
In this paper we simplify our computational methods by implementing a new
approach which regards expectation values  as a response to external fields. The inspiration to go from a ``measurement" to 
a ``response" based approach is derived from equilibrium quantum
mechanics. 

A major difficulty in the Keldysh approach, is that the measurement 
vertex is much more awkward to use than the ``response'' vertex.
For example, in the Larkin Ovchinikov basis the response vertex 
is diagonal  whereas the measurement vertex is 
off-diagonal ($-i\tau _{1}$). 
This leads to serious difficulties 
when dealing  with a conserved order parameter,
for double poles  which appear in the perturbation theory  can not
be absorbed using integration by parts. 
The same double poles are easily handled 
by regarding the expectation value of a quantity as the response to
its
conjugate field,  because 
the response vertex is proportional to the unit Keldysh matrix, and 
between two bare Green's 
functions it satisfies the relation
\[
{\mathcal{G}}\mathcal{G}= - \partial_{\omega}\mathcal{G}.
\]
This permits us
to eliminate  double poles in the perturbation theory using
integration by parts. It is also extremely useful in the development
of Ward identities. 

In equilibrium thermodynamics, 
we first calculate the Free energy, and then we take the
derivative to compute the magnetization
\[
M = - \frac{\partial{F}}{\partial B}\]
Can an analogous approach be used out-of-equilibrium? 

We now make a digression on the nature of the
non-equilibrium steady state. One of the key features of thermal
equilibrium, is the notion that the equilibrium steady state
does  not depend on its past history. In particular,  if we turn
on some interaction of strength $g$, ${H}_{I}= g h_{I}$, where $h_{I}$ 
is a dimensionless interaction  operator, 
and we couple internal degrees of freedom $\hat  A_{i}$ ($i=1,n$) 
to corresponding external fields $\lambda_{i}$, then the amount of
work done on the system 
\[
\Delta W = \int_{P} \langle h_{I} (t)\rangle  dg (t) +  \langle A_{i}
(t)\rangle d\lambda_{i} ,
\]
where we have used the summation convention on $i$,
does not depend on the detailed path $P$ over which the coupling constants
$(g,\lambda_{i})$ are adiabatically incremented 
to their final value. This implies that the ``curl'' defined by the 
following functional derivatives must vanish
\begin{eqnarray}\label{}
\frac{\delta \langle h_{I} (t)\rangle }{\delta \lambda _{i} (t')} 
-
\frac{\delta \langle A_{i} (t)\rangle }{\delta g (t')} &=& 0\cr
\frac{\delta \langle A_{j}(t)\rangle }{\delta \lambda _{i} (t')} -
\frac{\delta \langle A_{i} (t)\rangle }{\delta \lambda _{j}} &=& 0.
\end{eqnarray}
>From linear response theory, we can relate these functional derivatives
to the corresponding response functions, 
\begin{eqnarray}\label{}
\frac{\delta \langle h_{I} (t)\rangle }{\delta \lambda _{i} (t')} &=&
-i\langle [h_{I} (t),A_{i} (t')]\rangle \theta (t-t')\cr
\frac{\delta \langle A_{i} (t)\rangle }{\delta g (t')} &=&
-i\langle [A_{i} (t),h_{I} (t')]\rangle \theta (t-t')\cr
\frac{\delta \langle A_{j} (t)\rangle }{\delta \lambda _{i} (t')} &=&
-i\langle [A_{j} (t),A_{i} (t')]\rangle \theta (t-t')
\end{eqnarray}
from which it follows that
\begin{eqnarray}\label{}
-i\langle [h_{I} (1),A_{i} (2)]\rangle \theta (1-2)&=&
-i\langle [A_{i } (1), h_{I} (2)]\rangle \theta (1-2),
\cr
-i\langle [A_{j} (1),A_{i} (2)]\rangle \theta (1-2)&=&
-i\langle [A_{i } (1), A_{j} (2)]\rangle \theta (1-2).
\end{eqnarray}
These relations are the quantum mechanical counterpart of the famous
Onsager reciprocity relations\cite{Onsager:1931a,Onsager:1931b}.  They reflect the
microscopic reversibility of the equations of motion and the absence
of any  ``arrow of time'' in thermal equilibrium.

We now argue that Onsager's reciprocity relation will continue to hold for 
an important sub-class of variables in the non-equilibrium steady state.
For our discussion, we consider a quantum dot that is coupled to two
very large baths of electrons (``leads'') at different chemical potentials
$\mu_{L}$
and $\mu_{R}$ where $\mu_{L}> \mu_{R}$  The entire system is isolated. 
Suppose we connect the quantum dot to the leads
at time $t=0$, then we expect that after some equilibration time
$\tau _{1}$, which is basically independent of the size of the leads,
the system will arrive at a steady state where a current of electrons
flows from the left, to the right-hand lead. This steady state
will persist for a long time $\tau _{2} (L)$
until a substantial fraction of the additional electrons on the left
lead have flowed into the right lead.  Typically, the time $\tau _{2} (L)$
over which the steady state will persist, will diverge as
$L\rightarrow \infty $. The steady state value of some variable $\hat
A$ is then given by
\[
\langle A\rangle  = \lim_{L\rightarrow \infty }\langle A (t)\rangle 
\]
with the understanding that $\tau _{2} (L)>>t>>\tau _{1}$. 

The steady state involves a persistent current from left to right,
and evidently involves an ``arrow of time''. Nevertheless, we
expect that the nature of the steady state does not depend on the 
history of how it is arrived at. In particular, suppose we
the steady state is arrived by adiabatically turning on an interaction
$H_{I}= gh_{I}$ between the leads, and by coupling source terms
$\lambda_{j}$ to various quantities $A_{j}$ associated with the dot. 
Since the system is closed, when  we adiabatically change these variables 
we can define the amount of work done
in reaching the steady state
\[
\Delta W_{NE} = \int \langle h_{I} (t)\rangle  dg (t) +  \langle A_{i}
(t)\rangle d\lambda_{i} ,
\]
We emphasize that 
since the system is completely isolated, the amount of work done in
reaching the final state is literally the change in the total energy 
of the system: a completely well-defined 
quantity. 
\widf=0.9\columnwidth\\
\figa{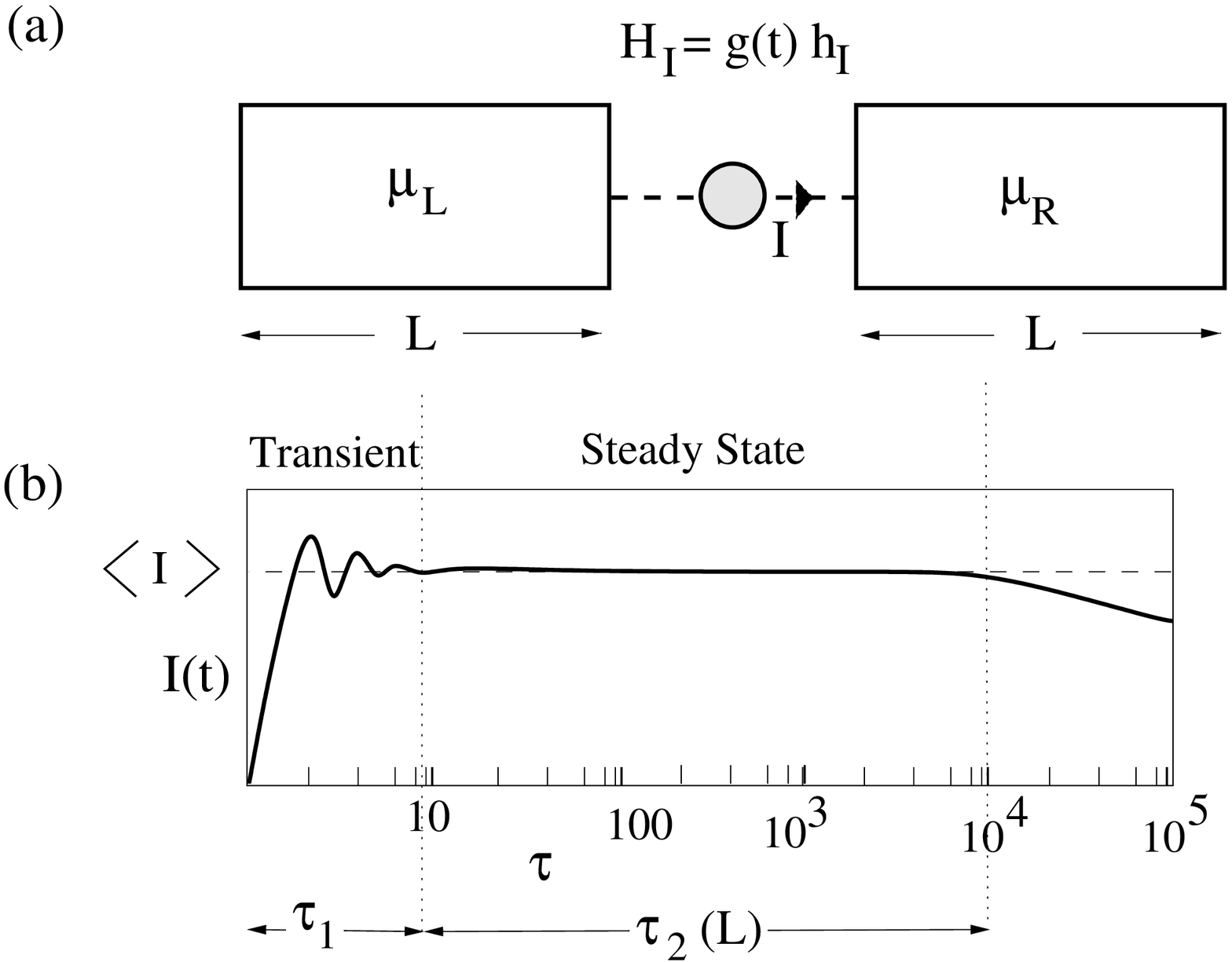}{Illustrating a closed quantum mechanical system 
in which the non-equilibrium steady state of a quantum dot can be studied.
(a) Quantum dot coupled to two isolated leads, initial in thermal
equilibrium at two different chemical potentials, 
via an
interaction term $H_{I}= g (t) h_{I}$. The coupling constant g (t)is
switched on at time $t=0$.
(b) After an initial transient of time $\tau _{1}$, the current
reaches a steady state value which only depends on the chemical
potential difference and the other fields (e.g the magnetic field)
applied to the dot.  This steady state persists for a large time
$\tau _{2} (L)$ which goes to infinity as the size of the leads is
sent to infinity. 
}
{newpic}\noindent 
If the steady state is independent of the path by which
it is arrive at, then it is not unreasonable to expect that 
there is a sub-class of variables $\{A_{i} \}$ for 
which the work $\Delta W_{NE}$ is path independent. 
If
this case, that clearly the Onsager relations given above
must extend to the non-equilibrium steady state for this subclass of
variables. Of course we do not expect
the reciprocity relation to extend to {\sl all} variables, as it does
in thermal equilibrium, because this would mean that the arrow of
time  is completely invisible. However, the idea 
that there
exists a subclass of ``conforming variables'' whose correlation
and reponse functions are completely symmetric in time, if true, could
be invaluable for studying the non-equilibrium steady state. 

This reasoning has motivated us to recently propose a ``Quantum Reciprocity
Conjecture''\cite{colemanmao:01} :
\begin{quote}
{\sl In the non equilibrium steady state, the set of quantum
mechanical observables
contains a non-trivial subset ${\mathcal{P}}$  of ``protected''
quantum observables ${\mathcal{P}}=\{a_{1},a_{2}\dots ,a_{n} \}$ whose
correlation functions in the steady state
are insensitive to the arrow of time, and which consequently satisfy a 
quantum mechanical analog of the Onsager reciprocity relations
\[
\langle [a (1),b (2)]\rangle = \langle [b (1),a (2)]\rangle , \qquad (a,b\
\in {\mathcal{P}}).
\]}
\end{quote}
Consider the retarded and advanced
Green functions between protected variables, 
\begin{eqnarray}\label{}
G^{( R,A)}_{ab} = \mp i \langle  [a (1),b (2) ]\rangle \theta_{\pm} (t_{2}-t_{1})
\end{eqnarray}
where $\theta _{\pm} (t)= \theta (\pm t)$. Since $a$ and $b$ are
hermitian, these are real
functions ($G^{R,A} (t)=[G^{R,A} (t)]^{*}$). The 
conjectured Onsager relations mean that in the steady state, 
they also satisfy
\begin{eqnarray}\label{}
G^{ R}_{ab} (t_{2}-t_{1})&=& G^{ A}_{ab}(t_{1}-t_{2}),\cr
G^{( R,A)}_{ab} (t_{2}-t_{1})&=&G^{( R,A)} _{ba}(t_{2}-t_{1}),
\end{eqnarray}
where the order of the subscripts and time variables is important.
If we write $G^{R}= G^{R*}$ in the first relation, and then
Fourier transform, 
we obtain the more familiar result
\[
G^{A}_{ab} (\omega )=G^{R}_{ab} (\omega)^{*}
\]
which means that the 
retarded and advanced Green functions of protected variables  share
the same spectral decomposition
\[
G^{{(R,A)}}_{ab} (\omega ) = \int \frac{dE}{\pi} \frac{1}{\omega
-E \pm i \delta }A_{ab} (E)
\]
where $A_{ab} (E)= \pm{\rm Im }[G^{(A,R)}_{ab} (E)]$.

Onsager used his reciprocity relations to propose that an
effective ``Free energy'' might exist in a non-equilibrium steady
state\cite{Onsager:1931a,Onsager:1931b}, an idea which has its origin
in early works by Lord Rayleigh\cite{Rayleigh}. Recently, Derrida,
Lebowitz
and Speer have shown that this idea holds  
rigorously in certain simplified non-equilibrium classical models
\cite{lebowitz}. 
Provided that the set of protected quantum variables includes the interaction
$h_{I}$, then it must follow that one can define an effective Free energy
from the virtual work done in reaching the steady state
\begin{eqnarray}\label{}
\Delta F[g,\{\lambda _{j} \}] &=& \int \langle h_{I} \rangle  dg  +  
\langle A_{i}
\rangle d\lambda_{i} \cr
&=&\int \frac{\langle H_{I}\rangle  }{g} 
dg  +  \langle A_{i}
\rangle d\lambda_{i} .
\end{eqnarray}
We have tested this conjecture on a simple, non-interacting
resonant level model\cite{colemanmao:01}, where 
the hybdization with the leads $H_{I}= gh_{I}$, the 
occupancy, magnetization and spin current through the dot
are found to satisfy mutual Onsager
reciprocity relations in the presence of a finite steady state charge current. 
The current operator does not lie within the class of protected variables,
nor indeed does  any operator which changes the balance of coupling
between the leads, or with a separate heat bath.   Nevertheless, the existence
of a finite subclass of protected variables has been confirmed in this
non-interacting  example.

We now apply the above arguments on the assumption that  the 
interactions inside the quantum dot are also protected variables. 
If this is the case, then interactions can be adiabatically 
turned on inside the quantum dot, whilst
preserving the reciprocity relations. 
This assumption enables us to extend the classic equilibrium
expression for the 
change in 
magnetization as a consequence of coupling to the leads, 
\[
\Delta M = -  \frac{\partial{}}{\partial B}
\int \frac{dg}{g}\langle {{H}}_{I}\rangle ^{( Eqn)}_{g}
\]
to the non-equilibrium situation
\[
\Delta {\cal M} = -  \frac{\partial{}}{\partial B}
\int \frac{dg}{g}\langle {{H}}_{I}\rangle ^{(NE)} _{g},
\]
This ``energetically defined'' magnetization 
quantity measures the response of the interaction
energy to the external field in a the non-equilibrium steady state
and it is an interesting quantity in its own right.
We shall now show
that this quantity
is directly related to the change in the non-equilibrium magnetization
due to interactions, 
\[
\Delta M_{NE} = \Delta {\cal M}.
\]
provided  Quantum Reciprocity holds between the interaction and the
magnetization. 

Consider a non-equilibrium system in the steady state. 
To be careful, let us 
focus on some variable in the Hamiltonian, call it $A$, which
is
conjugate to variable $\lambda  (t)$, so that the Hamiltonian has the form
$H= H_{o}+ {{H}}_{I}+ \lambda  (t)A (t)$, where we implicitly assume that
the interaction ${{H}}_{I}=
g h_{I}$ has strength g, which we ultimately set to unity.
Imagine turning on the interaction, the change in the A is
then given by
\[
\frac{\partial A (t_{o})}{\partial g} = -i \frac{1}{g}\int_{-\infty }^{t_{o}}
\langle [A (t_{o}), {{H}}_{I} (t')]\rangle dt'
\]
where the response function is to be evaluated at non-zero $g$. 
If quantum reciprocity holds, 
\[
-i\langle [A (t_{o}),H_{I} (t')]\rangle = 
-i\langle [H_{I} (t_{o}),A (t')]\rangle ,
\]
so that we can now put
\begin{eqnarray}\label{}
\frac{\partial A (t_{o})}{\partial g} = -i \frac{1}{g}\int_{-\infty }^{t_{o}}
\langle [{{H}}_{I} (t_{o}), A (t')]\rangle dt'
\end{eqnarray}
But we can simplify this expression by writing
\[
-i\langle [{{H}}_{I} (t_{o}), A (t')]\rangle\theta (t_{o}-t')
=  \frac{\delta \langle {{H}}_{I} (t_{o})\rangle }{\delta \lambda (t')}
\]
so that 
\begin{eqnarray}\label{}
\frac{\partial A (t_{o})}{\partial g} & =& -\frac{1}{g}\int_{-\infty }^{t_{o}}dt'
\frac{\delta \langle {{H}}_{I} (t_{o})\rangle }{\delta \lambda (t') }\cr
&=& \frac{1}{g}
\frac{\partial \langle {{H}}_{I} (t_{o})\rangle }{\partial \lambda
}.\end{eqnarray}
Finally, integrating over $g$, we have 
\[
\Delta A (t_{o}) = \frac{\partial}{\partial \lambda}\int_{0}^{1}\frac{dg'}{g'} \langle {{H}}_{I} (t_{o})\rangle _{g'}
\]

We will apply this expression to the magnetization in the
non-equilibrium Kondo model, making the
replacement $A\rightarrow M$ and  $\lambda\rightarrow -B$ and
$g\rightarrow J$ to obtain
\[
\Delta M = -  \frac{\partial{}}{\partial B}
\int \frac{dJ}{J}\langle {{H}}_{I}\rangle ^{( NE)}.
\]
This expression deserves some discussion.  In deriving it, 
we have assumed that the magnetization can relax to its new steady
state value : a very delicate assumption!

The magnetization is actually a  conserved quantity which commutes
with the full Hamiltonian $[M,H]=0$, and so strictly
speaking, by coupling the dot to the leads, we \underline{fail}
to relax the total magnetization. 
To avoid this paradox, we must 
surreptitiously introduce a regulated 
magnetization operator whose matrix elements lie between states
which are not degenerate, as in the case of a conserved operator, but 
whose energies are split by  an energy scale $s$. Such an operator
will 
consequently
relax on a time scale $\tau _{r}=\hbar /s$. It is important that this time-scale
is faster than
the adiabatic time scale $\tau _{a}\sim \hbar /\delta $
at which the interaction is switched on, where $\delta $ is the
regulator used inside the Green functions, i.e. 
\[
\tau _{r}=\hbar /s << \hbar /\delta = \tau _{a}.
\]
This is done in practice by weakly connecting the leads and the local moment
to a heat bath, e.g. by considering the local moment to 
weakly coupled to a spin chain.  Formally, at the end of the
calculation, we take $\delta \rightarrow 0$  at fixed $s$, then let
$s$ go to zero:
\[
M = \lim_{s\rarrow 0 }\lim_{\delta \rarrow 0} \langle M\rangle
_{s,\delta }
\]
Formally, a precisely similar regulation is required
in equilibrium. For example, when we consider 
the Pauli spin susceptibility, which is computed by taking the
spin susceptibility at a finite $q$ which is ultimately set to zero. 
\[
\chi _{Pauli}= \lim_{q\rarrow 0 }\lim_{\omega \rarrow 0} \chi
(\vec{q},\omega)
\]
Fortunately, 
we need only consider this process in the abstract, 
because the computation of the interaction energy in a field does not
present one with any divergent singularities. Difficulties
in earlier calculations associated with point-splitting are 
completely eliminated. 	

%
%
%

\section{Definition of Green Functions}

Let us denote by $t_+$ a time lying on the upper branch of the contour, and
by $t_-$ a time lying on the lower branch.  Then we may formally define four
Green's functions:
\bea
G^{++}(t,t') & \equiv & -i \langle T \psi(t_+) \psi^\dag(t'_+) \rangle; \nonumber \cr
G^{+-}(t,t') & \equiv & i \langle \psi^\dag(t'_+) \psi(t_-) \rangle; \nonumber \cr
G^{-+}(t,t') & \equiv & -i \langle \psi(t_-) \psi^\dag(t'_+) \rangle; \nonumber \cr
G^{--}(t,t') & \equiv & -i \langle {\tilde T} \psi(t_-) \psi^\dag(t'_-) \rangle. \label{gf1}
\eea
Here, $T$ is the usual time ordering operator, while ${\tilde T}$ is the
anti-time-ordering operator; we consider the operators $\psi$ to be
fermionic.  The Green's functions in (\ref{gf1}) are not all independent.  In fact,
there are only two independent Green's functions here.  This is most easily seen
by employing a transformation into the so-called ``Larkin-Ovchinnikov'' basis\cite{LO}.

Let us define the matrix ${\bf G}'$, whose entries are the four Green's functions of
(\ref{gf1}):
\be
{\bf G}' \equiv \pmatrix{G^{++} & G^{+-} \cr G_{-+} & G_{--} \cr}.
\ee
The transformation to the Larkin-Ovchinnikov basis is defined as
\be
{\bf G}' \to {\bf G} \equiv {\bf U} {\bf G}' {\bf V},
\ee
where the matrices ${\bf U}$ and ${\bf V}$ are given by
\be
{\bf U} = {1 \over \sqrt{2}} \pmatrix{1 & 1 \cr 1 & -1 \cr}, \qquad
{\bf V} = {1 \over \sqrt{2}} \pmatrix{1 & 1 \cr -1 & 1 \cr}.
\ee
The elements of this matrix Green-function correspond to the
Green-functions of the symmetric and antisymmetric fields
\begin{eqnarray}\label{}
\psi _{S} (t)&=& \frac{1}{\sqrt{2}}\left(\psi (t_{+})+\psi (t_{-})\right)\cr
\psi _{A} (t)&=& \frac{1}{\sqrt{2}}\left(\psi (t_{+})-\psi (t_{-})\right).
\end{eqnarray}
Defining 
\begin{equation}\label{}
\Psi (t)= \left(\matrix{\psi _{S}\cr \psi _{A}} \right), \qquad \bar
\Psi (t)= \psi \dg (t)\tau _{1} = ( \psi \dg _{A},\psi\dg _{S})
\end{equation}
then 
\begin{eqnarray}\label{}
{\bf G} (1,2) &=& -i \langle P \Psi (1) \bar \Psi (2)\rangle \cr 
&=&-i\langle P\left[
\left(\matrix{\psi_{S} (1)\cr \psi _{A} (1)} \right)
\otimes \left(\matrix{\psi\dg  _{A} (2)&\psi\dg  _{S} (2)} \right)
 \right]\rangle 
\end{eqnarray}
where $P$ denotes ``path-ordering'' of the fields along the  Keldysh contour.
It is easily shown that the  joint fluctuations in the antisymmetric
fields identically vanish, (Appendix A)
\begin{eqnarray}\label{}
G_{Y} (1,2)&=& -i\langle P \psi _{A} (1)\psi\dg  _{A} (2)\rangle =0,
\end{eqnarray}
and that the remaining matrix elements can be written as follows
\be
{\bf G} = \pmatrix{G^R & G^K \cr 0 & G^A \cr}, \label{loform}
\ee
where the physical Green's functions are given by
\bea
G^R(t,t') & = & -i \, \theta(t-t') \left\langle \left\lbrace \psi(t),\psi^\dag(t')
\right\rbrace \right\rangle; \nonumber \\
G^A(t,t') & = & i \, \theta(t'-t) \left\langle \left\lbrace \psi(t),\psi^\dag(t')
\right\rbrace \right\rangle; \nonumber \\
G^K(t,t') & = & -i \left\langle \left[ \psi(t),\psi^\dag(t') \right] \right\rangle. \label{physgf}
\eea
The first two Green's functions encode the dynamics whereas,
the third, $G^K$ (the Keldysh Green's function), encodes the evolution
of the particle distribution.  This, of course, is unnecessary in equilibrium, since
the particle distribution is known if the dynamic Green's functions are given.
The two indices of the matrix ${\bf G}$ will be called ``indices in Keldysh space''.

Let us now discuss the specific propagators for the quantum dot Kondo model.
One technical difficulty is that the Hamiltonian $H_{\rm I}$ contains,
the spin operator ${\bf S}$.  To furnish a Wick's theorem, 
it proves convenient to factorize ${\bf S}$ in terms 
of canonical creation and annihilation operators. 
This is the  ``Abrikosov
pseudo-fermion representation''\cite{Abrikosov:1965}
\bea
\vec{S} = f\dg _{\alpha }\left( \frac{\vec{\sigma }
}{2}\right)_{\alpha \beta }
f_{\beta }, \label{pseudof}
\eea
where $f\dg _{\alpha }$ creates a pseudo-fermion with spin component
$\alpha $. To provide a faithful representation of the spin $1/2$
operator, we must impose the additional constraint $n_{f}=1$. This is
done very conveniently by using the ``Fedatov-Popov trick''\cite{Popov0,Popov}, in which the 
 f-electron field in the past is equilibrated with a heat bath
at a complex chemical
potential $\mu=-i \pi T/2$. 
With this device, the contributions of doubly
occupied and empty states exactly cancel.  The occupancy of the
f-electrons in the distant past is 
\[
n_{f\sigma }=f (-\sigma B +i \pi T/2)= \frac{1}{e^{ -\sigma B/T +i \pi /2 }+1},
\]
where $f(x)$ is the Fermi function.
The magnetization of a free spin then recovers the Brillouin function
\[
M^{(0)} (T,B)= \sum_{s=\pm 1}\sigma n_{f\sigma } = \tanh (\frac{B}{T}).
\]
yielding a Curie susceptibility $\chi ^{0}= \partial
M/\partial B\vert_{B=0}= \frac{1}{T}$.

The diagrammatic elements corresponding to the bare Green's functions
are then
\begin{eqnarray}\label{}
\raisebox{-0.15truein}{\epsfig{figure=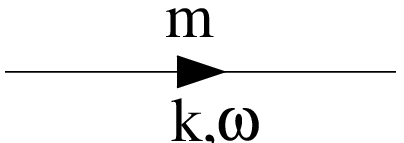,width=0.3\columnwidth}} 
\ =\ {\bf g}_{m} (k,\omega) 
\end{eqnarray}
for the conduction electron in the m-th lead and 
\begin{eqnarray}\label{}
\raisebox{-0.15truein}{\epsfig{figure=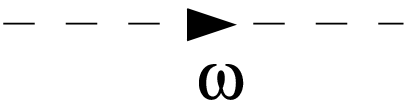,width=0.3\columnwidth}} 
\  =\  {\bf G}(\omega)
\end{eqnarray}
for the pseudofermion. Written out explicitly, 
\begin{eqnarray}\label{}
[{\bf g}_{ m} (k,\omega)&&]_{\sigma \sigma '} \equiv 
{\bf g}_{ m\sigma} (k,\omega)\delta _{\sigma \sigma '},\cr
\qquad \nonumber
\\
{\bf g}_{ m\sigma} (k,\omega)
=&&\pmatrix{g^{R}_{m\sigma } (k,\omega )
& g^{K}_{m\sigma } (k,\omega ) \cr
0 & g^{A}_{m\sigma } (k,\omega )\cr}\cr
\nonumber
\\
=&&\pmatrix{{1 \over \omega - \epsilon_{\bk \sigma} + i \delta}
& 2\pi i  h_{m}(\epsilon_{\bk\sigma }) \delta(\omega - \epsilon_{\bk \sigma}) \cr
0 & {1 \over \omega - \epsilon_{\bk \sigma} - i \delta} \cr},
\end{eqnarray}
where $
\epsilon_{{\bf k}\sigma }= 
\epsilon_{{\bf k}}-\sigma B 
$ and  
\begin{eqnarray}\label{}
[{\bf G}(\omega)&&]_{\sigma \sigma '} \equiv  
{\bf G}_{\sigma }(\omega)\delta _{\sigma \sigma '} \cr
\qquad \nonumber
\\
{\bf G}_{\sigma }(\omega)
=&&\pmatrix{G^{R}_{\sigma } (\omega )
& G^{K}_{\sigma } (\omega ) \cr
0 & G^{A}_{\sigma } (\omega )\cr}\cr
\nonumber
\\
 = && \pmatrix{{1 \over \omega
-\lambda _{\sigma }+ i \delta} & 
 2\pi i \, 
h_{\sigma }
\delta(\omega - \lambda _{\sigma }) 
 \cr
0 & {1 \over \omega -\lambda _{\sigma }- i \delta} \cr}.
\end{eqnarray}
where we have introduced the notation $\lambda _{\sigma }=
-\sigma B$. 
The functions $h_m(\epsilon)=h (\epsilon - \mu_{m})$  and
$ h_{\sigma  }  =h (\lambda _{\sigma }
+ \frac{i\pi T}{2})$ give the occupancies of 
the band and f-electron states, where 
\be
h(\epsilon ) \equiv 2 f(\epsilon ) - 1 = - \tanh \left( \epsilon 
\over 2T \right).
\ee
The quantity $\delta $ is the usual infinitesimal
regulator distinguishing the retarded and advanced Green's functions. 

\section{Feynman Rules}

In developing the Feynman rules we need to consider two types of vertices. The
``scattering vertex'' generated by contractions of terms
$H_{I}^{A}=H_{I} (t_{+})-H_{I}
(t_{-})$
in the expansion of the time
ordered exponential is denoted by
\begin{eqnarray}\label{response}
\raisebox{-0.4truein}{\epsfig{figure=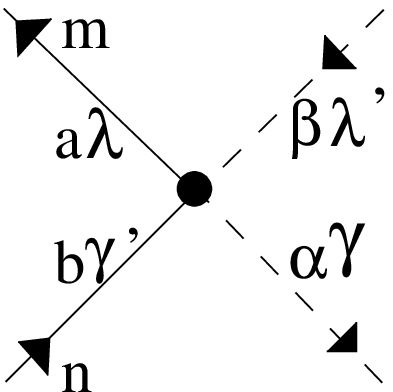,width=0.25\columnwidth}}
&\equiv
 \raisebox{-0.37truein}{\epsfig{figure=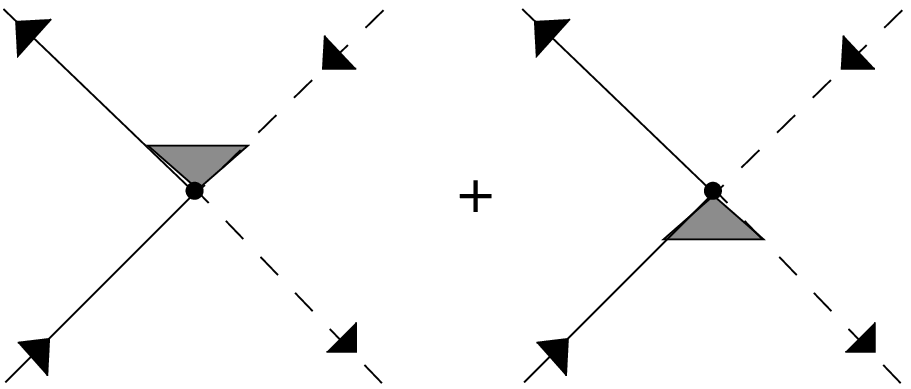,width=0.5\columnwidth}}
\cr\quad \nonumber
\\
= \frac{J_{mn}}{2}\vec{
\sigma }_{ab}\cdot \vec{\sigma }_{\alpha \beta }&
\left[ (\frac{i\btau_{\!1}}{2}
)_{\lambda \lambda '} \otimes \bone_{\gamma \gamma
'} + \bone _{\lambda \lambda '}\otimes (\frac{i\btau_{\!1}}{2}
)_{\gamma \gamma  '}\right].
\end{eqnarray}
(where  the factor 
``$i$'' derives from one factor of $-i$ in the time ordered, 
exponential, and two factors of $i$ transferred from external legs).
  Here, $\alpha ,\beta ,a, b$ refer to the spin
indices and $\lambda ,\lambda',
\gamma ,\gamma '$ to the Keldysh indices of the outgoing and
incoming propagators.
The factors in bold are the
matrices in Keldysh space that appear at the vertex.  
The shaded triangle denotes the vertex to which the Keldysh matrix
$\frac{i\btau_{\!1}}{2}$ is to be applied. 
The Keldysh matrix entering into the vertex is actually symmetric
under exchange of the Keldysh indices of the incoming, or outgoing
fermions, 
\begin{eqnarray}\label{}
(\btau_{\!1})_{\lambda \lambda '} \otimes \bone_{\gamma \gamma
'} + \bone _{\gamma \gamma '}\otimes (\btau_{\! 1})_{\lambda \lambda '}&=&\cr 
 (\btau_{\!1})_{\lambda \gamma '} \otimes \bone_{\gamma \lambda 
'}&+& \bone _{\gamma \lambda '}\otimes (\btau_{\! 1})_{\lambda \gamma '}
\end{eqnarray}
so that at each scattering vertex one has a choice as which pairs
of incoming and outgoing fermion lines one applies the Keldysh matrices
to. 

 Diagrammatically, 
\begin{eqnarray}\label{}
\raisebox{-0.25truein}{\epsfig{figure=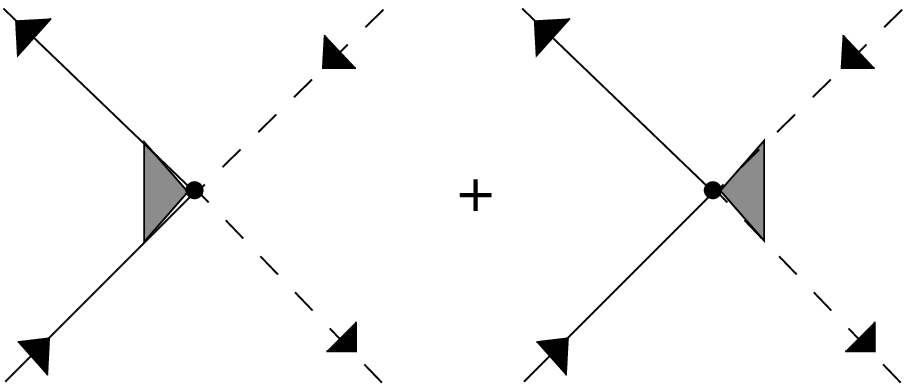,width=0.4\columnwidth}}
&\equiv&
\raisebox{-0.25truein}{\epsfig{figure=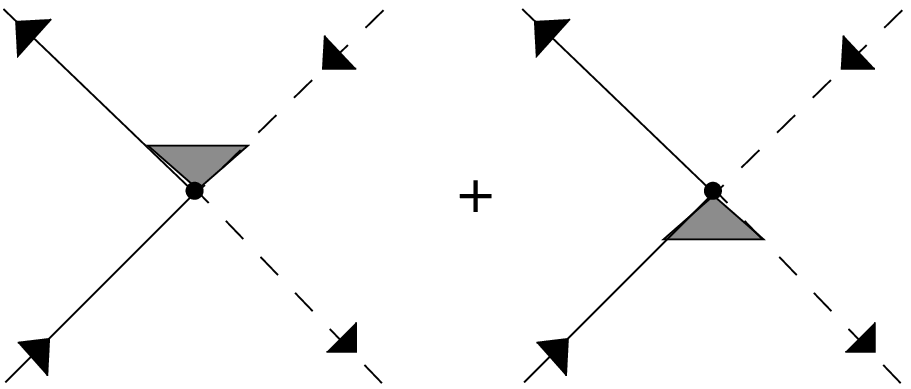,width=0.4\columnwidth}}
\end{eqnarray}

The ``measurement vertex'' generated by contractions with
$H_{I}^{S}=\frac{1}{2}[H_{I} (t_{+})+H_{I} (t_{-})]$ is denoted by
\begin{eqnarray}\label{measurement}
\raisebox{-0.4truein}{\qquad \epsfig{figure=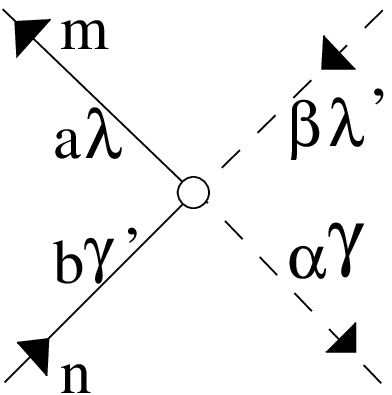,width=0.25\columnwidth}}
\quad \equiv & \raisebox{-0.4truein}{\epsfig{figure=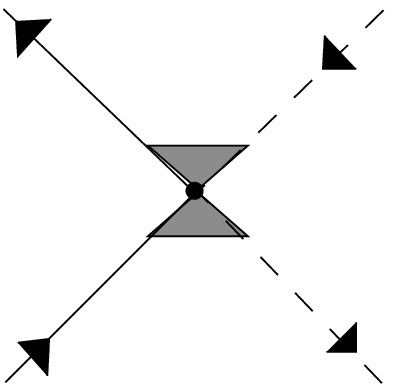,width=0.25\columnwidth}}
\qquad 
\cr
= \frac{J_{mn}}{2}\vec{
\sigma }_{ab}\cdot \vec{\sigma }_{\alpha \beta }
&\left[ (\frac{i\btau_{\!1}}{2}
)_{\lambda \lambda '} \otimes (\frac{i\btau_{\!1}}{2}
)_{\gamma
\gamma '} 
\right],
\end{eqnarray}
(where the factor ``$i^{2}= -1$'' from the $\tau_{1} $ matrices
is transferred from the contractions of two external propagators.)
Although the vertex $(\btau_{\!1})_{\lambda \lambda '} \otimes (\btau_{\! 1})_{\gamma
\gamma '} $ is not symmetric under interchange of incoming Keldysh
indices, one can make it symmetric by adding in the additional
``null'' term 
${\bf 1}_{\lambda \lambda'}\otimes {\bf 1}_{\gamma \gamma '}$.  The
expectation value of this vertex is zero, since it involves the
product of two operators that are antisymmetric between the upper and lower
Keldysh contour. 
By  introducing this operator, using the identity
\[
(\btau_{\!1})_{\lambda \lambda '} \otimes (\btau_{\! 1})_{\gamma
\gamma '} + \bone_{\lambda \lambda '} \otimes \bone_{\gamma \gamma '}=
(\btau_{\!1})_{\lambda \gamma '} \otimes (\btau_{\! 1})_{\gamma
\lambda '} + \bone_{\lambda \gamma '} \otimes \bone_{\gamma \lambda  '}
\]
and then removing the null operator on the right-hand side, 
one also has the freedom to interchange the
Keldysh indices on the measurement vertex, i.e. 
\begin{eqnarray}\label{}
\raisebox{-0.125\columnwidth}{\epsfig{figure=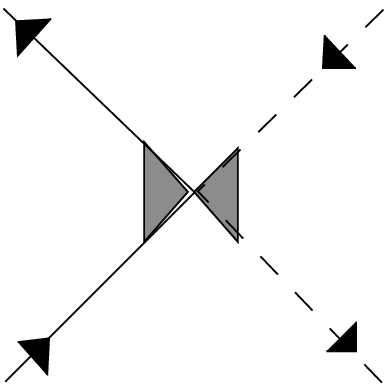,width=0.25\columnwidth}}
&\equiv&
\raisebox{-0.125\columnwidth}{\epsfig{figure=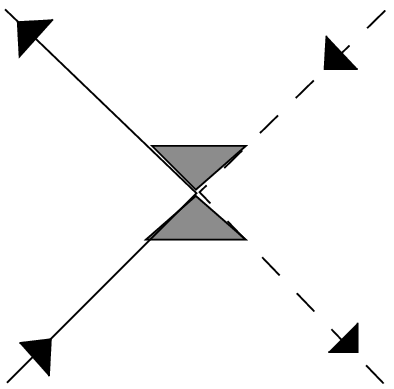,width=0.25\columnwidth}}
\end{eqnarray}

Our Feynman rules for the evaluation of the n-th order 
contribution to $\langle H_{I}\rangle $ are 
then 
\begin{itemize}

\item Construct a connected diagram with one measurement vertex
(\ref{measurement}) and $n$ scattering vertices (\ref{response})

\item Associate factors $ {\bf
g}_{m}$ and ${\bf G}(\omega )$ with the
conduction and pseudofermion propagators. 

\item Associate a factor $-1$ with each conduction electron and
pseudo- fermion loop. 

\item Carry out the trace over spin and Keldysh indices. 

\item Carry out the summation 
over internal momenta $\sum_{k}$. 

\item Carry out the integration over
loop frequencies $\int \frac{d\omega}{2\pi}$. 
Associate the convergence factor  $\cos \omega 0^{+}=
\frac{1}{2}\left( e^{i \omega 0^{+}}+e^{-i \omega 0^{+}} \right)$
with propagators which form closed loops. 

\item Associate a symmetry factor $1/n$ with 
diagrams with permutation symmetry group of dimension $n$.

\end{itemize} 

\section{Some useful diagrams}

\subsection{Vanishing Diagrams}

In constructing the Feynman diagrams, there are a number of
key diagrams which identically vanish. Any loop diagram which does
not contain an insertion of the Keldysh matrix $i\tau _{1}$
automatically vanishes, because it corresponds to the expectation
value of a single, or a product of antisymmetric densities.
For example,  the simple fermion loop  corresponds to the
equal time expectation value of the difference between fermion
number on the upper and lower Keldysh contour
\begin{eqnarray}\label{}
\raisebox{-0.06\columnwidth}{\epsfig{figure=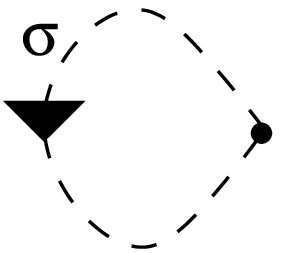,width=0.15\columnwidth}}
&=& -i 
\langle \bar f_{\sigma } (t) f_{\sigma } (t)\rangle \cr &=& -i \langle
n_{f\sigma }(t_{+})-
n_{f\sigma }(t_{-})
\rangle =0
\end{eqnarray}
In the frequency domain this result is obtained as follows
\begin{eqnarray}\label{}
\raisebox{-0.06\columnwidth}{\epsfig{figure=fig8d.eps,width=0.15\columnwidth}}
&=& -\int \frac{d\omega }{2\pi } \Tr [{\bf G}_{\sigma } (\omega )] \frac{1}{2}\left( e^{i \omega 0^{+}}+e^{-i \omega 0^{+}} \right)\cr
&=& -\int \frac{d\omega }{4\pi } \left[ G^{R}_{\sigma } (\omega
)e^{-i \omega 0^{+}}+G^{A}_{\sigma } (\omega ) e^{i \omega 0^{+}}
\right] 
\cr
&=&-\frac{1}{\pi } (-i\pi +i\pi )=0.
\end{eqnarray}
In this diagram, the first convergence factor catches the pole contribution
from $G_{A}$, whereas the second convergence factor catches an equal
yet opposite contribution from $G_{R}$. 
Two other examples of these cancellations are 
\begin{eqnarray}\label{}
\raisebox{-0.06\columnwidth}{\epsfig{figure=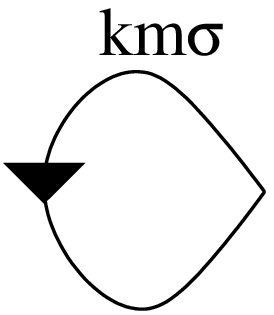,width=0.15\columnwidth}}
&=& -i \langle
n_{mk\sigma }(t_{+})-
n_{mk\sigma }(t_{-})
\rangle  =0.
\end{eqnarray}
and
\begin{eqnarray}\label{}
\raisebox{-0.06\columnwidth}{\epsfig{figure=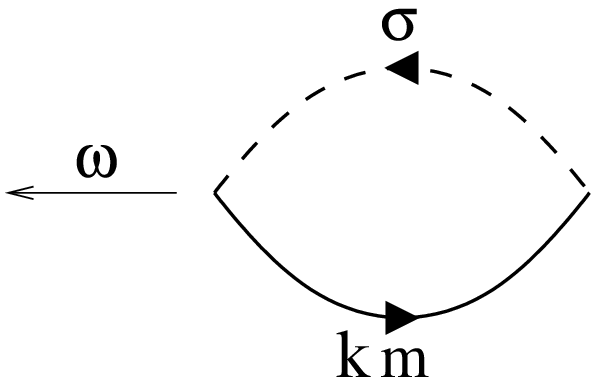,width=0.22\columnwidth}}
&=& -\int \frac{d\nu }{2\pi }\Tr \left[ {\bf G}_{\sigma } (\omega +\nu
){\bf g}_{m\sigma } (k,\nu  ) \right] \cr &=& 0.
\end{eqnarray}
where in this case, the cancellation does not require a convergence factor.

\subsection{Magnetization and Susceptibility}

The inclusion of the factor $-i\tau _{1}/2$ into fermion loops gives
the equilibrium 
expectation value of one particle quantities. For example,  the equilibrium
magnetization of the local moment is given by
\begin{eqnarray}\label{}
\raisebox{-0.06\columnwidth}{\epsfig{figure=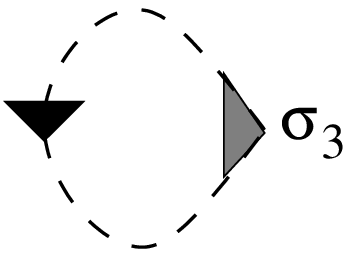,width=0.15\columnwidth}}
 &=& -1 \int \frac{d\omega }{2\pi } \Tr \left(\frac{i\tau
_{1}}{2}\sigma _{3}{\bf G} (\omega ) \right)\cr
&=& \sum_{\sigma }\sigma \int \frac{d\omega }{4\pi i } 
\left[ 2\pi i h_{\sigma }\delta (\omega -\lambda
_{\sigma })\right] \cr
&=& \sum _{\sigma }\frac{\sigma }{2}h (\lambda _{\sigma }-i\pi T/2)
\cr &=&
M_{o} (T,B )=
\tanh \left(\frac{B}{T} \right).
\end{eqnarray}
Using the identity
\begin{equation}\label{}
-\frac{\partial}{\partial B} {\bf G} (\omega )= 
{\bf G} (\omega )\sigma _{3}{\bf G} (\omega )  
\end{equation}
we can differentiate the above result to compute the equilibrium finite-field 
susceptibility of the local moment
\begin{eqnarray}\label{}
-\frac{\partial}{\partial B} \left[
\raisebox{-0.05\columnwidth}{\epsfig{figure=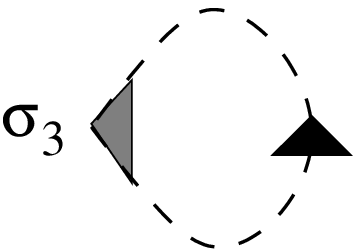,width=0.15\columnwidth}}
 \right]&=& 
\raisebox{-0.06\columnwidth}{\epsfig{figure=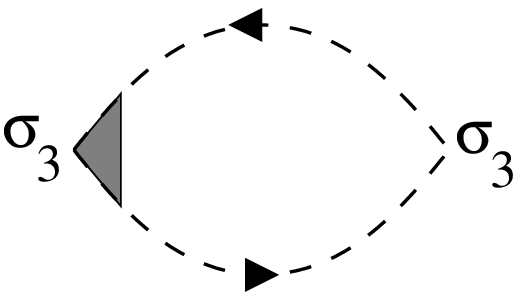,width=0.23\columnwidth}}\cr
 &=& -1 \int \frac{d\omega }{2\pi } \Tr \left(\frac{i\tau
_{1}}{2}\sigma _{3}{\bf G} (\omega ) \sigma _{3}{\bf G} (\omega ) 
\right)\cr
&=& -
\frac{\partial M_{o}}{\partial B} 
\end{eqnarray}

In a similar fashion the magnetization of the electrons in the m-th lead is given by
\begin{eqnarray}\label{}
\raisebox{-0.06\columnwidth}{\epsfig{figure=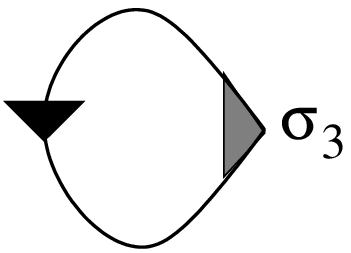,width=0.15\columnwidth}}
&=& -1 \sum_{k}\int \frac{d\omega }{2\pi } \Tr \left(\frac{i\tau
_{1}}{2}\sigma _{3}{\bf g}_{m} (k,\omega ) \right)\cr
&=& \sum_{k,\sigma } \frac{\sigma }{2}h_{m} (\epsilon _{k\sigma }) \cr
&=& 2 B \sum_{k } \left(-\frac{\partial f_{m}}{\partial \epsilon _{k}} \right)
= 2 \rho B
\end{eqnarray}
This is the only case where we have to explicitly consider the effect
of the magnetic field on the conduction electron lines.  For all
conduction electron propagators that do not loop back on themselves,
we can carry out the summation over momentum $k$ in advance of forming
the Feynman diagram.   We shall denote
\begin{eqnarray}\label{}
\raisebox{-0.15truein}{\epsfig{figure=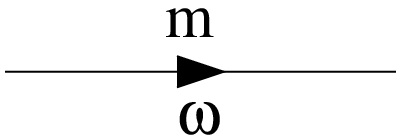,width=0.3\columnwidth}} 
\ =\sum_{k}  {\bf g}_{m} (k,\omega) = {\bf g}_{m} (\omega) \delta
_{\sigma \sigma '}
\end{eqnarray}
where 
\begin{eqnarray}\label{}
{\bf g}_{ m} (\omega)
=&&\pmatrix{g^{R}_{m } (\omega )
& g^{K}_{m} (\omega ) \cr
0 & g^{A}_{m} (\omega )\cr}\cr
\nonumber
\\
=&&\pmatrix{-i \pi \rho 
& 2\pi i \rho  h_{m}(\omega ) \cr
0 & i\pi \rho },
\end{eqnarray}
where we have used the large band-width approximation
\begin{equation}\label{}
\sum_{k}\frac{1}{\omega - \epsilon _{k\sigma }\pm i \delta } = \mp i
\pi \rho. 
\end{equation}
In this approximation,  the {\sl local} magnetic susceptibility of the
conduction electrons is given by
\begin{eqnarray}\label{locals}
\sum_{k,k'}\raisebox{-0.3truein}{\epsfig{figure=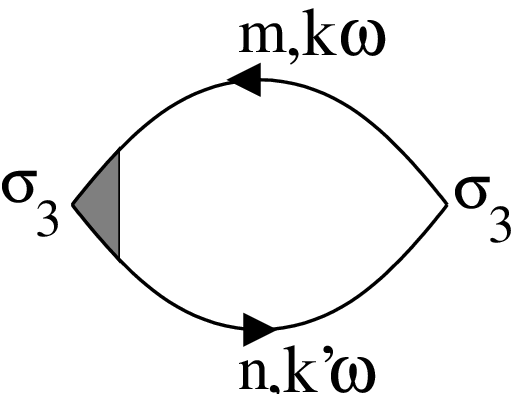,width=0.25\columnwidth}}
&=& -\int \frac{d\omega }{2\pi }\Tr \left[\frac{i\tau _{1}}{2}\sigma _{3}
{\bf g}_{m} (\omega )
\sigma _{3}{\bf g}_{n} (\omega )
 \right]\cr
&=& -\frac{\rho }{2}\int d\omega [h_{m} (\omega )-h_{n} (w)]
=0.
\end{eqnarray}
(a form of the Anderson-Clogston compensation theorem.)

\subsection{Kondo Polarization Bubbles}

Finally, for our calculations, it proves useful to be able to compute the
joint polarization bubbles between the f-electron and the conduction
electrons.  These bubbles control the renormalization of the interaction
between the local moment and the leads. 

There are three basic bubbles: 
\begin{eqnarray}\label{}
\raisebox{-0.27truein}{\epsfig{figure=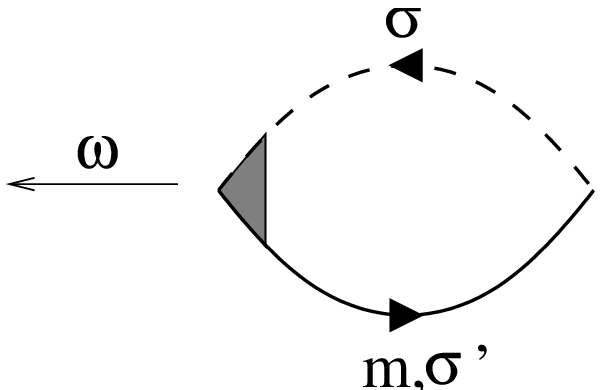,width=0.25\columnwidth}}
&=& \pi^{R}_{m\sigma } (\omega )\cr
\qquad \nonumber
\\
\raisebox{-0.27truein}{\epsfig{figure=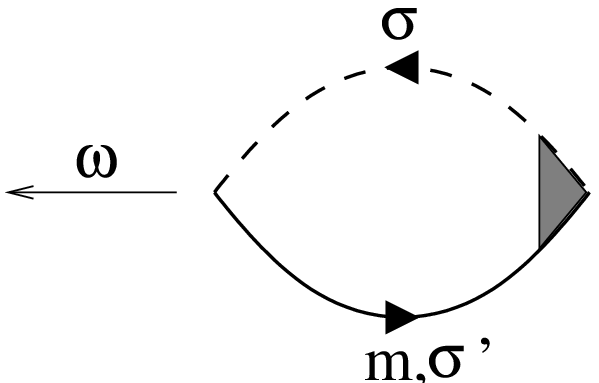,width=0.25\columnwidth}}
&=& \pi^{A}_{m\sigma } (\omega )= \pi^{R}_{m\sigma } (\omega )^{*}\cr
\qquad \nonumber
\\
\raisebox{-0.25truein}{\epsfig{figure=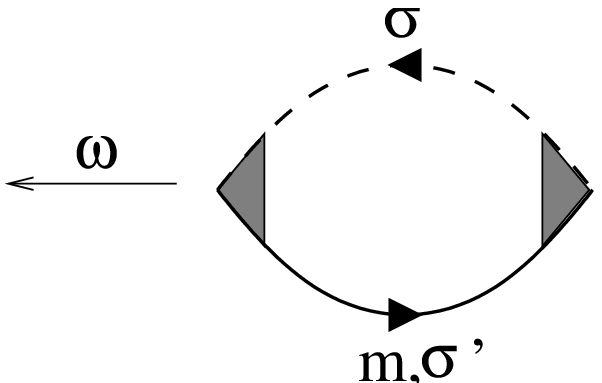,width=0.25\columnwidth}}
&=& \pi^{K}_{m\sigma } (\omega )\cr
\qquad \nonumber
\end{eqnarray}
Note that the spin index $\sigma $ on each bubble refers to the spin
component of the pseudo-fermion, and that each bubble is 
independent of the spin component ($\sigma
'$) 
of the conduction
electron propagator. 
The calculation of the first bubble proceeds as follows
\begin{eqnarray}\label{pir}
\raisebox{-0.2truein}{\epsfig{figure=fig8g.eps,width=0.2\columnwidth}}
&=&
-\int \frac{d\nu }{2\pi }\Tr \left[\frac{i\tau _{1}}{2}{\bf G}_{\sigma }
(\omega +\nu )
{\bf g}_{m} (\nu  )
 \right]\cr
&=& \int \frac{d\nu }{4\pi i}
\biggl[
G^{R}_{\sigma }
{g}^{K}_{m} 
+G^{K}_{\sigma }
{ g}^{A}_{m} 
 \biggr]_{\omega +\nu ,\nu }\cr
&=&\frac{1}{2}\sum _{k'}
\left(\frac{
h_{m} (\epsilon _{k'\sigma '})-h _{\sigma }
}{\omega +\epsilon
_{k'\sigma '}
-\lambda _{\sigma }+i\delta } \right).
\end{eqnarray}
In the large large band-width approximation, this becomes
\begin{eqnarray}\label{}
\pi_{m\sigma }^{R} (\omega ) = 
\frac{\rho }{2}\int d\nu 
\left(\frac{
h (\nu-\mu _{m} )
-
h _{\sigma }
)
}{\omega +\nu
-\lambda _{\sigma }+i\delta } \right).
\end{eqnarray}
Note how the field dependence of
the conduction electron propagator has been absorbed in the large band-width
limit by a small shift in the conduction electron energies 
$\epsilon _{k\sigma '} \rarrow \epsilon _{k}$.
By making use of the identity 
\begin{equation}\label{ident}
\frac{1}{2}\int dx 
\frac{h (x )-h _{\sigma }}{x -\epsilon + i\Delta  } 
=\psi \left ({\textstyle \frac{1}{2}-\frac{\epsilon -i\Delta  }{2 \pi iT}}\right)- 
\ln \left( {\textstyle 
\frac{De^{-\frac{i\pi}{2} h_{\sigma}}}{2\pi T }}\right) 
\end{equation}
where $\Delta >0$, and $\psi (z) = d\ln \Gamma (z)/dz$ is the digamma function,
we obtain
\begin{eqnarray}\label{hard1}
\pi^{R}_{\sigma } (\omega ) &=& 
\rho \left[
\psi \left (
{\textstyle \frac{1}{2}+
\frac{\omega -\lambda _{\sigma
}+\mu _{m}}{2 \pi iT}}
\right)
- 
\ln \left( {\textstyle \frac{De^{-\frac{i\pi}{2} h_{\sigma }}}{2\pi T }}\right)
\right]
,
\cr
\pi^{A}_{\sigma } (\omega ) &=& 
\rho \left[
\psi \left (
{\textstyle \frac{1}{2}-
\frac{\omega -\lambda _{\sigma
}+\mu _{m}}{2 \pi iT}}
\right)
- 
\ln \left( {\textstyle \frac{De^{\frac{i\pi}{2} h_{\sigma }}}{2\pi T }}\right)
\right]
.
\end{eqnarray}
The Keldysh loop is computed as follows
\begin{eqnarray}\label{pik}
\raisebox{-0.2truein}{\epsfig{figure=fig8i.eps,width=0.2\columnwidth}}
&=&-\int \frac{d\nu }{2\pi }\Tr \left[\frac{i\tau _{1}}{2}
{\bf G}_{\sigma }
(\omega +\nu )\frac{i\tau _{1}}{2}
{\bf g}_{m} (\nu  )
 \right]\cr
&=& \int \frac{d\nu }{8\pi }
\biggl[
G^{R}_{\sigma }
{ g}^{A}_{m}
+G^{A}_{\sigma }
{ g}^{R}_{m} 
+G^{K}_{\sigma }
{ g}^{K}_{m} 
 \biggr]_{\omega +\nu ,\nu }\cr
\nonumber
\\
&=&\frac{\pi }{2}\sum_{k}\left[1-h_{\sigma
}h_{m} (
\epsilon _{k\sigma '})\right]
\delta (\lambda _{\sigma }-\omega  -\epsilon
_{k\sigma '})
.
\end{eqnarray}
so that in the large band-width limit
\begin{eqnarray}\label{hard2}
\pi^{K}_{\sigma } (\omega )= \frac{\pi \rho }{2}\left[1-h_{\sigma }h_{m} (\lambda
_{\sigma }-\omega )
\right].
\end{eqnarray}

\section{Detailed Calculation}

We now discuss the evaluation of the interaction energy, 
up to second-order perturbuation theory. The corresponding Feynman
diagrams
are 
\begin{eqnarray}\label{}
\int_{0}^{J} \frac{dJ'}{J'} \langle {\mathcal H}_{I}\rangle _{J',H}
&=&
\raisebox{-0.14truein}{\epsfig{figure=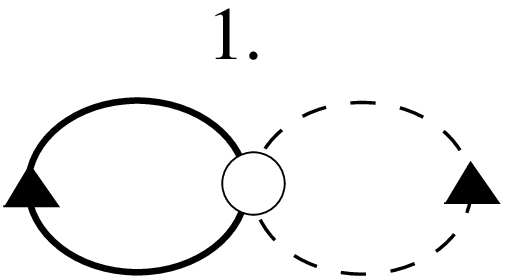,width=0.3\columnwidth}}
\cr
\qquad \nonumber
\\
+\frac{1}{2}\left[
\phantom{\raisebox{-0.3truein}{\epsfig{figure=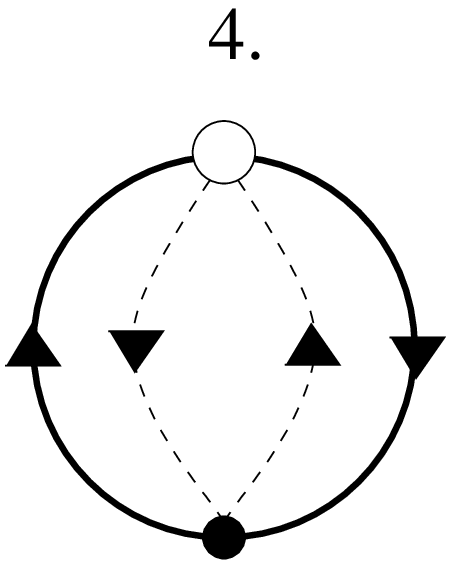,width=0.2\columnwidth}}
}\hskip -0.5truein
\raisebox{-0.1truein}{\epsfig{figure=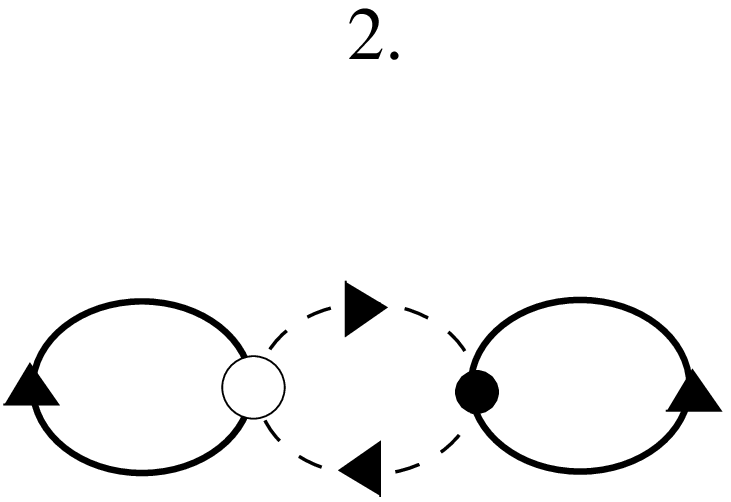,width=0.2\columnwidth}}
\right . &+& \left .
\raisebox{-0.1truein}{\epsfig{figure=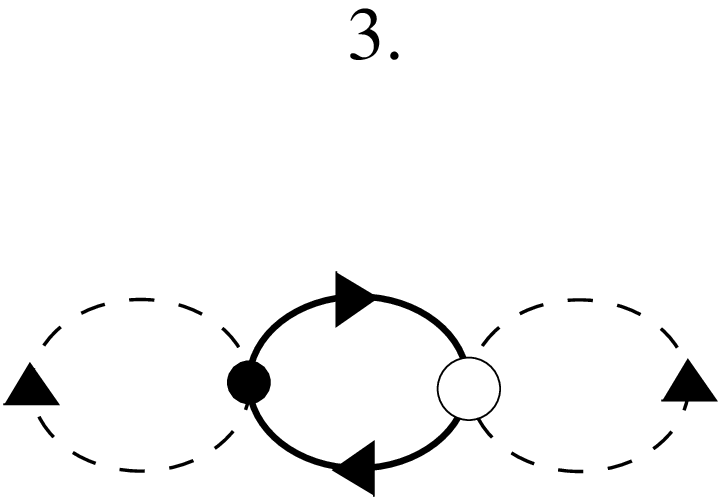,width=0.2\columnwidth}}
+\raisebox{-0.3truein}{\epsfig{figure=fig1b.eps,width=0.2\columnwidth}}
\right]\cr
\qquad \nonumber
\\
\cr
= \Delta F_{(1)}+\Delta F_{( 2)}&+&\Delta F_{( 3)}+\Delta F_{( 4)}
\nonumber
\end{eqnarray}
where an additional factor of $1/2$ in the last two terms 
comes from the coupling constant
integration.

In evaluating these diagrams, it is useful to divide the interaction
vertex into an ``Ising'' and ``x-y'' component, writing
\begin{equation}\label{spins}
 \frac{J_{mn}}{2}\vec{
\sigma }_{ab}\cdot \vec{\sigma }_{\alpha \beta }=
J_{mn}\left[ 
\frac{1}{2}\sigma^{3}_{ab}{\sigma ^{3}}_{\alpha \beta }+
\sigma ^{+}_{ab}\sigma ^{-}_{\alpha \beta }+
\sigma ^{-}_{ab}\sigma ^{+}_{\alpha \beta }
\right]
\end{equation}
where $\sigma ^{\pm}= \frac{1}{2} (\sigma ^{x}\pm i\sigma ^{y})$ are
the raising and lowering operators.  Notice that the amplitude
for the spin flip terms is twice that for the Ising terms.
Since the propagators are
diagonal in the spin indices, diagrams (1-3) only involve
the Ising component in the interaction. 

Taking account of the
vanishing diagrams, we may compute the first three diagrams as follows:
\begin{eqnarray}\label{details}
\boxy{fig1a.eps}{0.05}{0.3}&=&\boxy{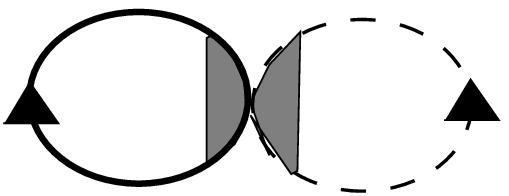}{0.05}{0.3}\cr
&=& \sum_{m}\frac{J_{mm}}{2} (2\rho B)M_{o} \cr
&=& (\bar J_{R}+\bar J_{L}) B M_{o}
\cr
\frac{1}{2}\left. \boxy{fig1c.eps}{0.04}{0.3}\right.
&=&\frac{1}{2}\left. \boxy{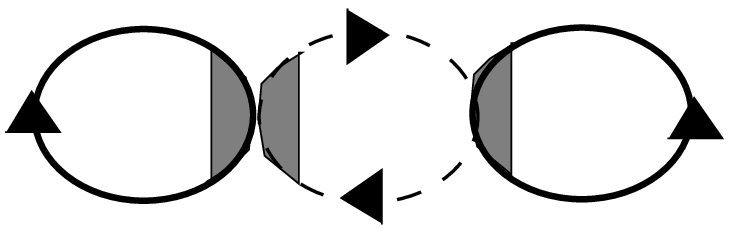}{0.04}{0.3}\right. \cr
&=& \frac{1}{2}\sum_{m,n}\frac{1}{4}J_{mm}J_{nn} (2\rho B)^{2}
\left(-\frac{\partial M_{o}}{\partial B} \right)\cr
&=& -\frac{1}{2} (\bar J_{R}+\bar J_{L})^{2}B^{2}\left(\frac{\partial M_{o}}{\partial B} \right)
\cr
\frac{1}{2}\boxy{fig1d.eps}{0.04}{0.3}&=&\frac{1}{2}\boxy{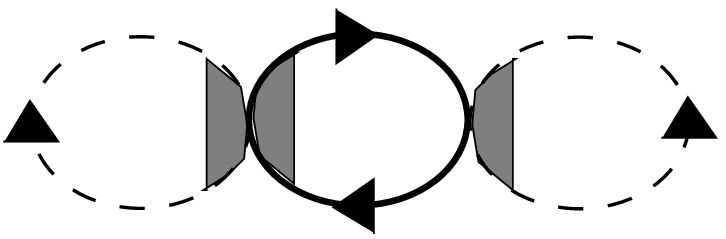}{0.04}{0.3}=0
\end{eqnarray}
where the vanishing of the local conduction electron susceptibility 
(\ref{locals}) causes the last term to vanish.

The sum of the first three terms in the ``free energy''
is then 
\begin{eqnarray}\label{}
 \Delta F_{(1)}+\Delta F_{( 2)}+\Delta F_{( 3)} =
\bar J BM_{o}- \frac{1}{2}(\bar JB)^{2}\frac{\partial M_{o}}{\partial B} ,
\end{eqnarray}
where $M_{o}= \tanh (\frac{B}{T})$ is the magnetization of the free
impurity, and we have put $\bar J_{R}+\bar J_{L}= \bar J$.  
We recognize these terms as the leading order expansion of 
the Free energy of a spin in a Weiss field $B_{eff}= ( 1 - J)B$:
\begin{eqnarray}\label{123}
F^{*}&=& - T \ln \left[ 2 \cosh (B/T ( 1 - \bar J) )
\right]
\cr
&=& - T \ln \left[ 2 \cosh (B/T )\right]
+ 
\bar J BM_{o}- \frac{1}{2}(\bar JB)^{2}\frac{\partial M_{o}}{\partial
B}+\dots  .
\end{eqnarray}
The Ising corrections to the energy are clearly independent of the
voltage between the leads. 

The last diagram in the interaction energy can be expanded as follows: 
\begin{eqnarray}\label{}
\frac{1}{2}\boxy{fig1b.eps}{0.095}{0.2}&=&
\frac{1}{2}\left[ \boxy{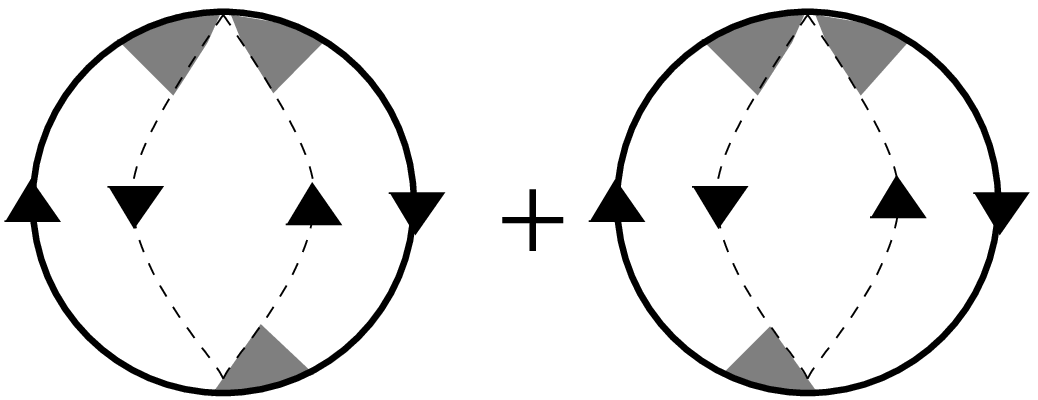}{0.07}{0.4}\right]
\end{eqnarray}
These expressions can be divided into Ising (non-spin-flip) 
and x-y (spin flip) components, $\Delta F_{(4)}= 
\Delta F^{z}_{(4)}+\Delta F^{\pm}_{(4)}
$, where 
\begin{eqnarray}\label{}
\Delta F^{z}_{(4)}&=& \frac{1}{2}\left[ \boxy{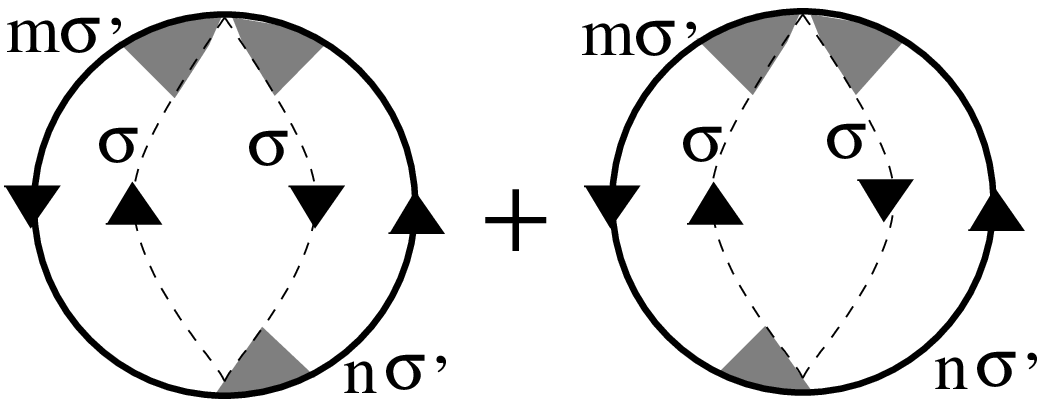}{0.07}{0.4}\right]\cr
&=& \sum_{m,n,\sigma }\frac{J_{mn}^{2}}{4}\int
\frac{d\omega }{2\pi }
\left(\pi^{R}_{m\sigma } 
\pi^{K}_{n\sigma } +
\pi^{K}_{m\sigma } 
\pi^{A}_{n\sigma } 
 \right)_{\omega }\cr
\Delta F^{\pm}_{(4)}&=& \frac{1}{2}\left[ \boxy{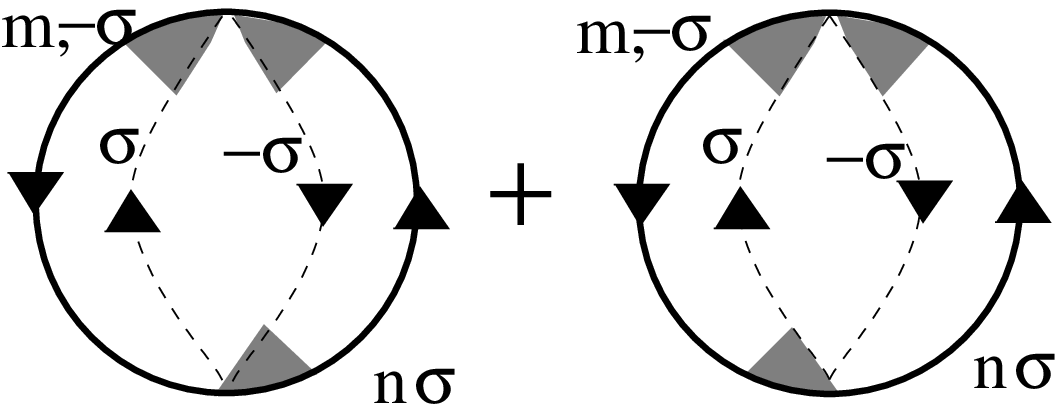}{0.07}{0.4}\right]\cr
&=& \frac{1}{2}\sum_{m,n,\sigma }J_{mn}^{2}\int
\frac{d\omega }{2\pi }
\left(\pi^{R}_{m\sigma } 
\pi^{K}_{n,-\sigma } +
\pi^{K}_{m\sigma }
\pi^{A}_{n,-\sigma } 
 \right)_{\omega }\cr
\end{eqnarray}
( The $\frac{1}{2}$ prefactor in the first equation
has been cancelled by the sum over the spin of the conduction electron
lines, and the factor of $(J_{mn})^{2}/4$ is derived from the square 
of the Ising scattering amplitude.  In the second equation, the spin
of the conduction electrons is set by the spin of the f-electrons, so 
the $\frac{1}{2}$ prefactor remains, however, the amplitude for
two spin flips is $( J_{mn})^{2}$, without a factor of $1/4$. )
Inserting the discrete expressions (\ref{pir} ) and (\ref{pik})
we obtain
\begin{eqnarray}\label{discrete}
\Delta F^{z}_{(4)}&=& {\rm Re }\sum_{mn\sigma }\frac{J_{mn}^{2}}{16}
\left\{ \frac{\textstyle ( 1-h_{\sigma }h_{nk}) (
 h_{\sigma }
-
h_{mk'}
)
}{\epsilon _{k }-\epsilon _{k'}} \right\},
\cr 
\Delta F^{\pm}_{(4)}&=&{\rm Re }\sum_{mn\sigma }\frac{J_{mn}^{2}}{8}
\left\{ \frac{\textstyle ( 1-h_{-\sigma }h_{nk}) (
h_{\sigma }
-
h_{mk'} 
)}{\epsilon _{k}-\epsilon _{k'
}-2\sigma B }
\right\}.
\end{eqnarray}
where we have denoted $h_{mk}\equiv h_{m} (\epsilon _{k})$.
Following earlier discussion, the field dependence of the band-electron
energies has been dropped in these expressions.

As a simple check on these results, 
consider the zero-temperature, equilibrium 
limit of (\ref{discrete} ). In this
limit, one can replace $h_{m} (\epsilon_{k'} )\rarrow -\sgn \epsilon _{k'}$,
$h_{\sigma }=\sigma $. The first term vanishes and the second term
becomes
\begin{eqnarray}\label{}
\Delta F^{\pm}_{(4)}&=&{\rm Re }\sum_{m,n,\sigma }\frac{J_{mn}^{2}}{8}
\left\{ \frac{( 1-\sigma \sgn (\epsilon _{k})) (
\sigma 
+\sgn (\epsilon_{k'})
)}{\epsilon _{k}-\epsilon _{k'
}-2\sigma B }\right\}  \cr
&=& J^{2 }\sum_{\epsilon _{k}<0, \epsilon _{k'}>0}\frac{1}{\epsilon_{k} -\epsilon_{k'}-2B}
\end{eqnarray}
which is recognized as the second-order correction to the ground-state
energy due to quantum spin fluctuations. 

To proceed further, it is convenient to introduce the notation
\begin{eqnarray}\label{}
S_{mn\sigma } (x,B)&=& Re\left[ 
{
2 \pi i \ln  \tilde{\Gamma}
\left(x+2\sigma B +\mu _{m}-\mu _{n}
 \right)}
\right. 
\cr &-& 
\left. 
{( x+2\sigma B +\mu _{m}-\mu _{n}) \ln \left(\frac{D}{2\pi T} \right)
}
 \right]\cr
\tilde \Gamma (x)&=& \Gamma \left(\frac{1}{2}+\frac{x}{2
\pi i} \right)
\end{eqnarray}
which has the property that 
\[
{\mathrm Re} \pi ^{R}_{m\sigma } (x-\mu _{n})= \rho 
S'_{mn\sigma } (x,B)
\]
where the prime denotes derivative w.r.t. x. With this notation, using
(\ref{hard1}) and (\ref{hard2}) we obtain
\begin{eqnarray}\label{xy}
\Delta F^{\pm}_{(4)}&=&  \sum_{m,n } 
\frac{\bar J_{mn}^{2} }{4}
\int d\omega S'_{mn\sigma } (\omega,B )(1-\sigma M_{o}h (\omega ))
\cr\Delta F^{z}_{(4)}&=&  \sum_{m,n } 
\frac{\bar J_{mn}^{2} }{8}
\int d\omega S'_{mn\sigma } (\omega ,0)(1+\sigma M_{o}h (\omega ))
\end{eqnarray}
where we have replaced ${\rm Re}h_{-\sigma }= -\sigma M_{0}$. 
Integrating this expression by parts, and using the fact that
$S_{mn\sigma } (\pm D)= \mp D$ we obtain
\begin{eqnarray}\label{xy3}
\Delta F^{\pm}_{(4)}
&=&  -\sum_{m,n,\sigma } \bar J_{mn}^{2} \left[D
-\sum_{\sigma } \sigma  \frac{M_{o}}{4}
\int d\omega 
h' (\omega )
S_{m n \sigma } (\omega)
 \right]\cr
\Delta F^{z}_{(4)}
&=& - \sum_{m,n,\sigma } \bar J_{mn}^{2} \frac{D}{2}
\end{eqnarray}
where the residual integral in $\Delta F^{z}_{(4)}$ vanishes  under the
spin summation. 
It is convenient to separate the logarithmic term away from
the integral over $S_{mn\sigma } (\omega )$, writing
\begin{eqnarray}\label{124}
J_{mn}^{2}\sum_{\sigma } &\sigma & \frac{M_{o}}{4}
\int d\omega 
[-h' (\omega )]
S_{m n \sigma } (\omega) \cr
&=& -2J_{mn}^{2}B M_{o}\left[
\ln \frac{D}{2\pi T}- \phi \left(\frac{2B}{T}, \frac{\mu _{m}-\mu _{n}}{T}\right)  \right]
\end{eqnarray}
where 
\begin{eqnarray}\label{125}
\phi (b,v)= \int \frac{dx}{(2\cosh \frac{x}{2})^{2}}\frac{1}{4b}\hskip -0.1in
\sum_{\sigma,\gamma =\pm 1} \hskip -0.1in{2\pi i \sigma 
\ln \tilde{\Gamma}(x + \sigma b+ \gamma
v )}
\end{eqnarray}

Combining the results (\ref{xy3}), (\ref{123}), (\ref{124}) and (\ref{125})
our final result for the change in the Free energy  $\Delta F=
\int \frac{dJ}{J }\langle
H_{I}\rangle $ is then 
\begin{eqnarray}\label{}
\Delta F &=&  -T (\ln \left[ 2 \cosh (B (1-\bar J) /T )\right] -F_{eqn}
- \frac{3}{2}\bar J^{2 }D\cr
&+&4\bar J_{RL}^{2}M_{o}B\left[ \ln \frac{D}{2\pi T} - \phi
\left(\frac{2B}{T},\frac{eV}{T} \right)
 \right]\cr
&+&2(\bar J_{R}^{2}+\bar J_{L}^{2})
M_{o}B\left[ \ln \frac{D}{2\pi T} - \phi
\left(\frac{2B}{T},0 \right)
 \right]\cr
&+& O (\bar J^{3}).
\end{eqnarray}
Notice that the appearance of $F_{eqn}$ on the right hand side of this expression does not rely
on any assumption about the limiting value of the Free- energy or
magnetization in the limit $\bar J\rarrow 0$. 
The change in the magnetization is then given by 
\begin{eqnarray}\label{}
\Delta M &=&\left(  (1-\bar  J)
\tanh \left[\frac{( 1- \bar J)B}{T}\right]
-
\tanh \left[\frac{ B}{T}\right]
\right)
\cr
&-&4\bar J_{RL}^{2}\frac{\partial}{\partial B}\left[ BM_{o}\left( 
 \ln \frac{D}{2\pi T} - \phi
\left(\frac{2B}{T},\frac{eV}{T} \right)
\right) \right]\cr
&-&2(\bar J_{R}^{2}+\bar J_{L}^{2})
\frac{\partial}{\partial B}\left[ BM_{o}\left( 
\ln \frac{D}{2\pi T} - \phi
\left(\frac{2B}{T},0 \right)
\right)\right]\cr
&+& O (\bar J^{3}).
\end{eqnarray}
Again, the appearance of the equilibrium magnetization on the
right-hand side of this expression has
nothing to do with the value of the magnetization in the limit $\bar r
J\rarrow 0$. With our method, we are unable to comment on the magnetization
in the limit $\bar J\rarrow 0$. However, the above expression 
is very strongly suggestive that the limit $\bar J \rarrow 0$ returns
to the
equilibrium expression for the magnetization, in which case, 
\begin{eqnarray}\label{}
M &=& 
(1-\bar J)\tanh \left[\frac{( 1- \bar J)B}{T}\right]
\cr
&-&4\bar J_{RL}^{2}\frac{\partial}{\partial B}\left[ BM_{o}\left( 
 \ln \frac{D}{2\pi T} - \phi
\left(\frac{2B}{T},\frac{eV}{T} \right)
\right) \right]\cr
&-&2(\bar J_{R}^{2}+\bar J_{L}^{2})
\frac{\partial}{\partial B}\left[ BM_{o}\left( 
\ln \frac{D}{2\pi T} - \phi
\left(\frac{2B}{T},0 \right)
\right)\right]\cr
&+& O (\bar J^{3}).
\end{eqnarray}

There are two special limits of the above result.  In the limit
$T\rarrow 0$, we may take advantage of the asymptotic form 
\begin{eqnarray}\label{}
B\left[\ln  \frac{D}{2\pi T}\right. &-&\left.\phi \left(\frac{2B}{T},\frac{V}{T}
\right)\right]\longrightarrow 
\cr
&-&\frac{1}{2}\sum_{\gamma =\pm}\left(2B +\gamma V
\right)\ln \left[\frac{\vert 2B +\gamma V\vert }{2\pi D e} \right]
\end{eqnarray}
so that in this limit 
\begin{eqnarray}\label{hunk2a}
M &=& \left[1 - (\bar  J_{R}+\bar J_{L}) -  2(\bar J_{R}^{2} +\bar J_{L}^{2} 
)\ln \left(
\frac{D}{2B}\right)\right.\cr &-&
\left. (4 \bar J_{RL}^{2})\ln \left(
\frac{D}{\sqrt{\vert (eV)^{2}-(2B)^{2}
\vert}
}\right)\right]
\dots 
\end{eqnarray}

In the zero field limit $B\rightarrow 0$, $\phi (b,v)\rightarrow \phi (0,v)= \Phi (v)-1+\gamma 
$, where 
\begin{eqnarray}\label{}
\Phi (v)&=&
\hbox{Re} \int_{-\infty }^\infty 
\frac{dx}{4 \cosh^2 \bigl(\frac{x}{2}\bigr)}\left[
\tilde{\psi } (v+x)-\tilde{\psi}(x) \right]
\end{eqnarray}
and $\tilde{\psi}(x)=\psi ( \frac{1}{2} + i \frac{x}{2\pi} )$.
and we have used the result
\[
\int_{-\infty }^\infty 
\frac{dx}{4 \cosh^2 \bigl(\frac{x}{2}\bigr)}
\tilde{\psi } (x) = - (1+\gamma )
\]
where $\gamma= 0.5772\dots = - \psi (1)$ is the  Euler constant. 
In this limit, the magnetization takes the form 
\begin{eqnarray}\label{suscept}
M &=& \frac{B}{T}
\left\{
 1 - 2(\bar J_{R} +\bar J_{L}) -
4\left(
   \bar J_L^2 + \bar J_R^2
\right)
\ln \left(
       \frac{De^{\frac{3}{4}+\gamma}}{2\pi T 
}
    \right) 
\right. 
\cr
&&\left.
- 8|\bar J_{LR}|^2 \left[ \ln
 \left(
\frac{De^{\frac{3}{4}+\gamma}}{ 2\pi T 
} 
\right)
- \Phi  \left( \frac{V}{T} \right)
\right] 
\right\}.
\end{eqnarray}

%

%
%

\section{Discussion}

We should like to end with  a discussion of some of the more
controversial aspects of our work. There are two general topics
that deserve discussion:

\begin{itemize}

\item [*]  Does the magnetization of the quantum dot
revert smoothly to the equilibrium value $M_{o}=\tanh
\left(\frac{B}{T} \right)$ when the coupling to the leads is reduced
to zero? 

\item [*]  The issue of quantum coherence, and whether  a current
   through the quantum dot dephases the Kondo effect.

\end{itemize}

Both issues are areas where our current work differs in important ways
with parallel work carried out by others in the field.

\subsubsection{Does the magnetization have a perturbative expansion? }

The first of the above items brings our discussion into 
contact up with the issue of whether a
perturbative treatment of the magnetization in the quantum dot is
valid out-of-equilibrium.  
Although
strictly speaking, our approach, based on the Onsager reciprocity relations,
can not derive the leading term in the magnetization, the appearance
of a well-defined perturbation expansion, and the very form of our results
\begin{equation}\label{}
\Delta M =  \tanh (B (1-\bar J) /T) -\tanh (B /T) 
+ O (\bar J^{2})
\end{equation}
suggests 
that the leading order term in the magnetization is just the Brillouin
function $\tanh (B /T)$. 

Recent work by Parcollet and Hooley (PH)\cite{PH}, Rosch, Passke,
Kroha nd Wolfle \cite{Roschnew} and unpublished work by
Kaminski  and Glazman (KG)\cite{kaminskii2}
casts doubt on the viability of a perturbative approach.  PH
compute the local magnetization of a quantum dot by computing the
steady-state Green function of the local moment, and they conclude
that in the limit where the coupling of the local moment to the
leads becomes infinitesimal, the magnetization does not revert to 
its thermal value, but instead acquires the limiting value
\begin{eqnarray}\label{}
\tilde{M}_{PH} (T,B,V)&=& M_{o} f (\frac{2B}{T},\frac{V}{T})\cr
f (b,v)&=& \frac{\varphi (b)(1+\theta )}{\frac{1}{2 }\left(
\varphi ( b+v)+\varphi (b-v)\right) +
\theta \varphi (b)
}
\end{eqnarray}
where 
\[
\varphi (x)= \frac{x}{\tanh \frac{x}{2}}
\]
and
$\theta = \frac{J_{R}^{2}+J_{L}^{2}}{2 J_{RL}^{2}}$. A similar
expression was derived by KG using a master equation approach. 
Results in the recent paper by Rosch et al. also reduce to the same
form. 
This expression can not be expanded perturbatively in the
coupling constants, and if correct, casts serious doubt in the whole
viability of a perturbative approach to the non-equilibrium Kondo model.

The KGPH result assumes that the relevant magnetization of the quantum dot
is the polarization of the local dot spin, whereas we have argued that
because the spin of the quantum dot partially delocalizes into the leads,
one must consider the global magnetization. 
This leads to a fundamentally different 
philosophy about the way a quantum dot equilibrates its magnetization.
Clearly, the local magnetization is not a conserved quantity,
and it can be relaxed by its coupling to the leads. 
By examining this problem, PH find 
a non-thermal
polarization of the magnetic moment of the dot
in the limit of zero coupling to the leads. 
We argue conversely, that if we focus on the {\sl total magnetization}
then the coupling to the leads is not sufficient  to relax to the
appropriate steady state magnetization. 
In this case, a weak coupling between
the system and its surroundings is always required to relax the conserved
total magnetization, which is otherwise conserved,  even when the 
the dot and the leads are coupled. Let us call the coupling between
the
dot an its surroundings $s$. 
It is reasonable to suppose that 
physically reasonable results will be obtained 
so long as $s$ is
small compared with all physical energy scales in the quantum dot, such as 
the renormalized Kondo temperature, but large compared with the adiabatic
scale $\delta$ which governs the rate at which interactions are turned
on:
\[
\hbox{Physical energy scales, $(T_{K}^{*}, eV, \dots )$}>> s >> \delta 
\]

In  KGPH approach, in the limit of zero coupling to the leads, the
local moment of the quantum dot is left in a non-thermal polarization.
These results are most striking in the limit of zero
temperature, when the polarization of the quantum dot is given by
\[
\tilde{M}_{PH}=\mu_{B} \times \left\{
\begin{array}{lr}
\frac{2B (1+\theta )}{eV/\mu_{B}+ 2\theta B}&\qquad (\mu_{B}B<eV/2)\cr
1&\qquad (\mu_{B}B> eV/2)
\end{array}
 \right.
\]
for $k_{B}T<<2\mu_{B }B<< eV$, where we have restored $V\rightarrow
eV$ and $B\rightarrow \mu_{B}B$. 
To preserve this non-thermal distribution, the
spin must remain decoupled from any other thermal bath. In other words,
the PH result corresponds to the case where the coupling to an external
thermal bath is zero, i.e $s=0$.  

To get a feeling for the 
consequences of this result, 
consider
a quantum dot at a bias voltage of $100 mV$ with $\theta =1$.
According to the PH result, the dot would have a magnetization $M\sim
0.1\mu _{B}$ even in a field of 50 Tesla. 
In a real experiment,
-this magnetization
would relax back to one Bohr magneton, due to coupling to an external
bath. In practice, such a coupling is \underline{always} present, and 
can not be neglected when considering the expansion around
the limit $\bar  J =0$.   Just as in equilibrium, providing these couplings
are small compared  with the relevant energy scales of the system- in this
case the renormalized Kondo temperature- then a dynamical calculation of
the magnetization must include them. 

It is difficult to reconcile the 
proposed failure of a perturbative expansion for the
magnetization out-of-equilibrium  with the existence of 
a perturbative expansion for the current 
and the total interaction energy of the dot. The 
current and interaction energy of the quantum dot appear to have
a perfectly regular perturbation expansion. 
In the steady state, then the magnetization 
is determined from differential of the 
coupling constant integrated interaction
energy:
\begin{eqnarray}\label{}
\Delta M &=& -\frac{\partial}{\partial B} \int_{0}^ {J}
\frac{dJ'}{J'}\langle H_{I}\rangle. 
\end{eqnarray}
While the Onsager reciprocity relation does not hold for charge
current and interaction energy, we are able to relate the current 
to the derivative of the interaction energy with respect to the
interaction energy\cite{colemanmao:01}, 
\begin{eqnarray}
I &=& \left.-\frac{\partial}{\partial A}\right|_{A=0}
\langle H_{I}\rangle .
\end{eqnarray}
where the above quantities are to be calculated by including
a vector potential $A$ into the scattering potential (but not into the
measurement vertex of $H_{I}$), by replacing
\[
J_{RL}\rarrow J_{RL}e^{{i\frac{eA}{\hbar}}}, \qquad 
J_{LR}\rarrow J_{LR}e^{{-i\frac{eA}{\hbar}}}.
\]
Indeed, when we include these changes into our calculation of $H_{I}$, 
we do recover the established second order expression for the
current through the quantum dot (Appendix D)
\[
I =\frac{3e^{2}}{h} (\bar J_{RL})^{2 } V.
\]
Thus a single energy functional appears to successfully yield both
the magnetization and the current.  From this point-of-view,
a fundamental failure of perturbation theory would have to manifest
itself in the interaction energy. The smooth perturbative expansion
of this quantity again suggests that the problems encountered by PH
in their calculation are one of equilibration, rather than  a break
down of the perturbation theory at high temperatures or fields. 

>From the arguments of this section, it appears
that the proposed failure of the perturbative
expansion is tied up with two issues- the appropriate
definition of the Kondo magnetization, and the 
the issue of how the
magnetization relaxes. 
If indeed, as we have argued, the total magnetization is the important
variable for characterizing the Kondo effect, 
then a coupling  to an external bath is necessary in any
dynamic treatment of the total 
magnetization. It would be interesting in
further work to examine whether such a coupling
does dynamically relax the total magnetization to the 
energetically  defined  magnetization $\cal  M$. 

\subsubsection{Decoherence and Ward Identities}

Let us now turn to the second issue: that of decoherence.  The general
issue, whether a non-equilibrium quantum system loses its phase
coherence is of great future importance, and will have bearing 
on non-equilibrium problems such as quantum computation. 
An issue here, fundamentally, is the question of whether 
driven non-equilibrium quantum
systems can develop phase coherent structure, or not.

In our quantum dot problem,
we have used the Onsager reciprocity relations to argue that the 
energetically defined magnetization
\[
{\mathcal{M}}= M_{eqn}- \frac{\partial }{\partial B}\int_{0}^ {J}
\frac{dJ'}{J'}\langle H_{I}\rangle 
\]
is identically equal to the steady state magnetization of the quantum
dot, 
\begin{equation}\label{}
{\mathcal{M}}\equiv M
\end{equation}
We are able to clearly demonstrate a perturbative expansion for
$\mathcal{M}$. 
Clearly 
even if the identity between the energetic and the physical
magnetization were to fail, 
the perturbative 
expansion of $\mathcal{M}$
can still be used as a criterion for whether the Kondo effect
enters a strong coupling regime at high voltages.   
The zero temperature, low field  result
\begin{equation}\label{}
{\mathcal{M}} = \left[1 - J\rho -  (J\rho )^{2} \ln \left(
\frac{D^{2}}{2B
\vert (eV) \vert}
\right)  \dots \right]
\end{equation}
shows that the {\sl leading} Kondo logs associated with intralead processes survive to arbitrarily high voltage. 

If decoherence is to prevent this divergence,
at large voltages, then presumably, the above
logs in $\cal  M$ would have to be cut-off by the decoherence rate
$\tau^{-1} \sim I/e$.  This would require that inside the logarithms,
we replace
\[
2B\rightarrow  \hbox{max}(2B, \hbar \tau^{-1}).
\]
For weak coupling to the leads, 
$\hbar \tau ^{-1}\sim \rho J^{2}
eV$ is the perturbative expansion for the current through the dot. 
Such cut-off effects \underline{will} clearly appear if we extend
the diagrammatics by dressing the fermion propagators using the Dyson equation
\begin{eqnarray}\label{}
{\bf \tilde{G} }^{-1} (\omega ) = 
{\bf {G} }_{0}^{-1} (\omega ) - 
\ \ \boxy{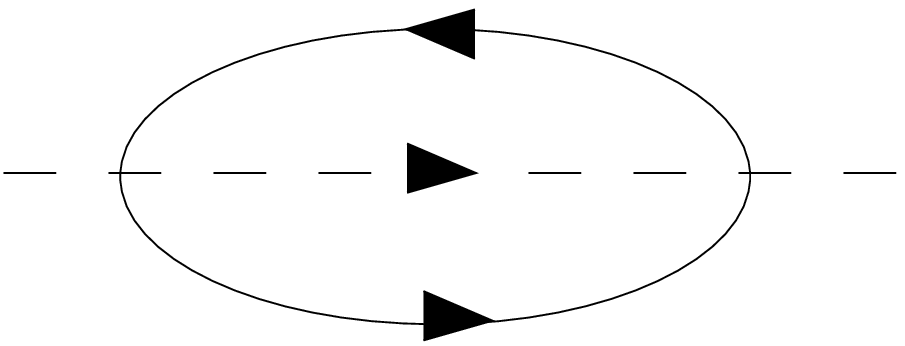}{0.03}{0.2}
\end{eqnarray}
The main effect of the self-energy insertions into the fermion lines
is to introduce a wavefunction renormalization of the
magnetization spin vertex
\begin{eqnarray}\label{}
( Z-1)g\mu_{B} =\  \ \ \boxy{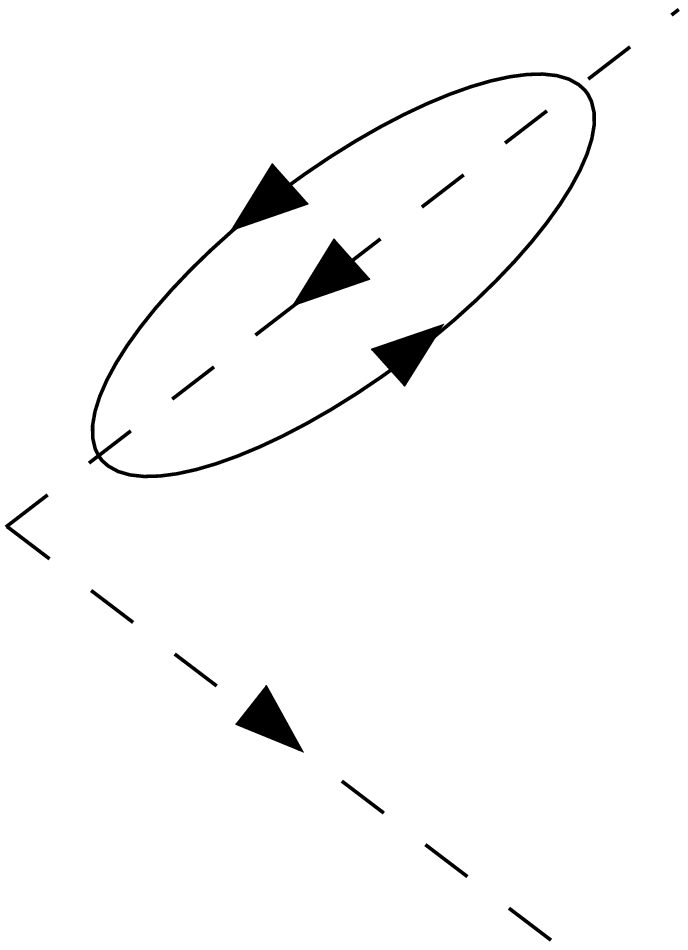}{0.06}{0.1}\ \ 
  \sim \left. \frac{\partial \Sigma (\omega )}{\partial \omega }\right
\vert _{\omega =0}g\mu _{B}
\end{eqnarray}
However, in any controlled systematic treatment of the diagrammatics,
it is important to also take into account the corresponding renormalization
of the magnetization vertex
\begin{eqnarray}\label{}
\boxy{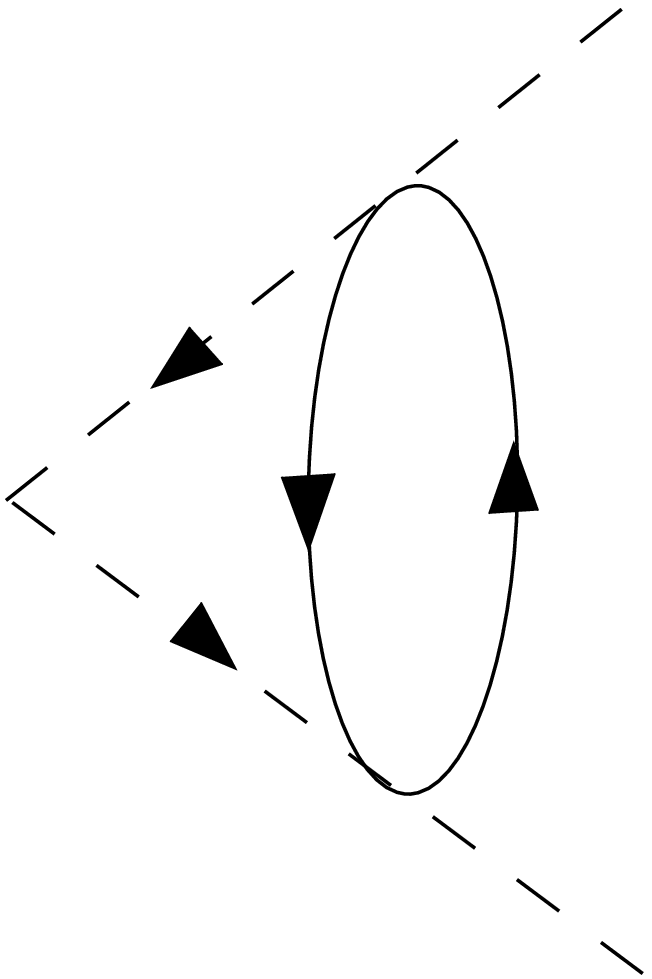}{0.06}{0.1}\ \ 
= \Gamma (\omega =0) g\mu_{B} 
\end{eqnarray}
Both of these diagrams are generated at the same time when one differentiates the
interaction energy with respect to the external magnetic field. 
Now the important point is that these two vertices are related by a Ward
identity associated with the conservation of total spin.  
In equilibrium, this leads to a Ward identity between 
the spin self-energy and the spin vertex
\[
\Gamma_{\omega=0}= \left. -\frac{\partial \Sigma (\omega )}{\partial \omega }\right\vert _{\omega =0}
\]
leading to the precise cancellation of these two terms.
\begin{eqnarray}\label{}
\boxy{disc2.eps}{0.07}{0.1}\ \ + 
\boxy{disc3.eps}{0.07}{0.1}\ \ 
= g\mu_{B} 
\left( \frac{\partial \Sigma (\omega )}{\partial \omega }+ \Gamma 
\right)_{\omega =0}
=0
\end{eqnarray}
This cancellation has important physical consequences. For example,
this is the reason why the renormlized ``meso-spin''
$S^{*} (\Lambda )$ of the sort discussed in the
introduction
\[
\vec{S}^{*}= e^{ {\cal  S}}\vec{S} e^{-{\cal  S}}.
\]
preserves a zero relaxantion rate, whereas the bare spin $\vec{S}$ relaxes
at a rate $\tau ^{-1}\sim J^{2}k_{B }T$ at high temperatures.
The point is, that the  magnetization of the local spin $M_{local}$ is
rigorously given by 
\[
M_{local} = n_{\uparrow} - n_{\downarrow }
\]
and this quantity therefore does \underline{not } contain any vertex
corrections.
By contrast, the total magnetization entering into the uniform
susceptibility of a quantum dot \underline{does} contain the
above vertex. From this we learn that the rapid fluctuations of the
local spin $\vec S$ described by the self-energy $\Sigma$ are also responsible
for the renormalization of the meso-spin $\vec{S}^{*} (\Lambda )$ and its
delocalization into the leads.  This is why these two effects cancel.
These considerations lead us to doubt any conclusions about the behavior
of the D.C. biased quantum dot which resum the self-energy corrections
to the pseudo-fermion lines without taking the corresponding vertex
corrections into account. Any calculations based on the 
``non-crossing approximation'' which assume that these vertices can be
ignored will lose these subtle cancellations. 

It is useful to consider what will happen to the leading logarithm
in the magnetic susceptibility if we continue to fourth order in the
computation.  The divergence we have found
\begin{equation}\label{}
\Delta \chi = - \frac{4 (\bar J_{R}^{2}+\bar J_{L}^{2})}{T}\ln
\left(\frac{De^{1+\gamma }}{2\pi T} \right)
\end{equation}
corresponds to the diagram
\begin{eqnarray}\label{}
\Delta M^{(2)} = \boxy{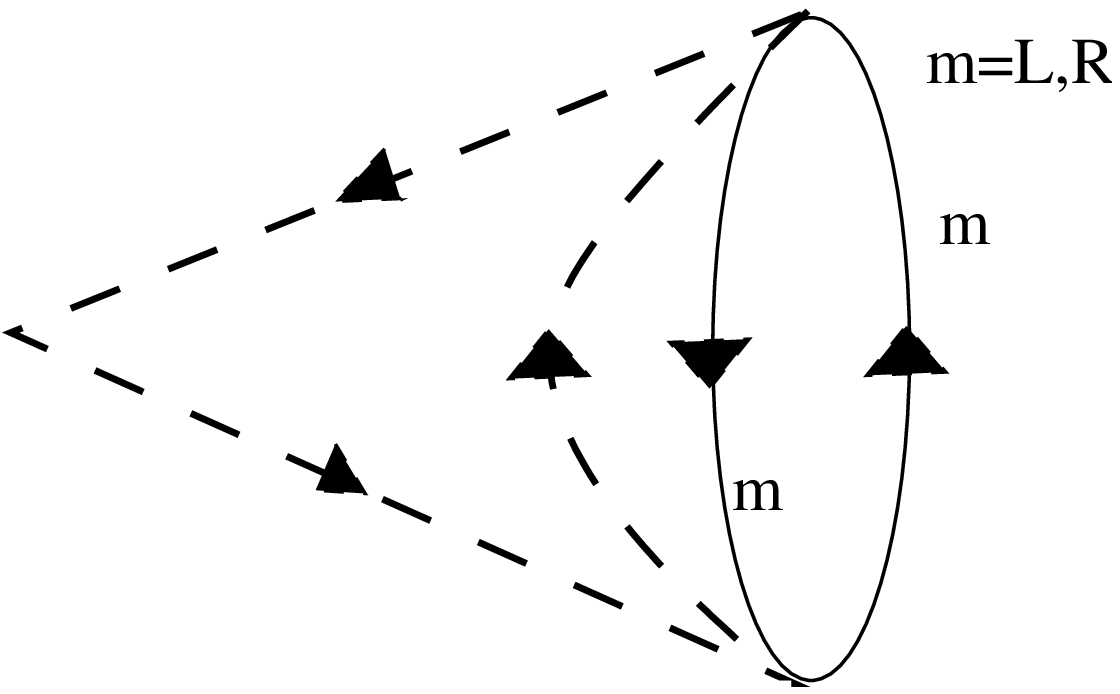}{0.08}{0.3}\ \ 
\end{eqnarray}
where the logarithm derives from the f-conduction loops. 
When we go to fourth order, we dress the internal f-lines
to produce the following three diagrams
\begin{eqnarray}\label{}
\Delta M_{\Sigma }^{(4)} = \boxy{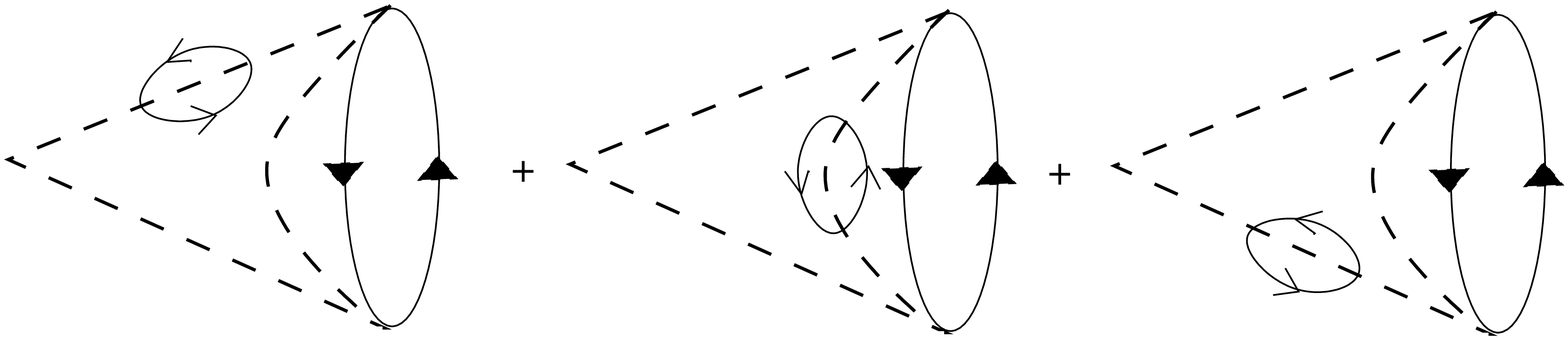}{0.075}{0.7}\ \ 
\end{eqnarray}
The renormalization effects of this 
diagram will tend to cut-off the logarithmic divergence.
However to the same order we also produce vertex counterterms
\begin{eqnarray}\label{}
\Delta M_{\Gamma }^{(4)} = \boxy{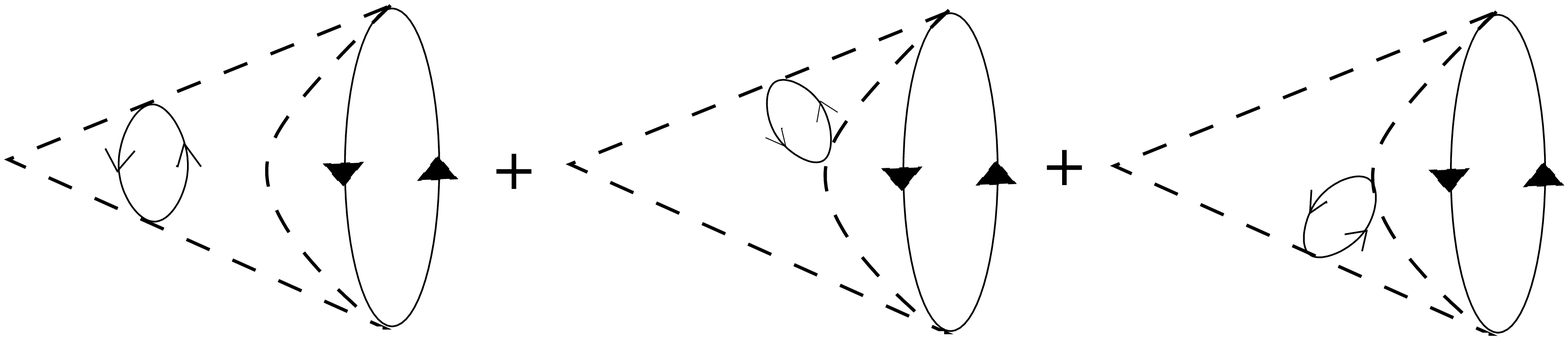}{0.075}{0.7}\ \ 
\end{eqnarray}
These vertex corrections correspond physically to the fact that
the meso-spin $\vec{S}^{*} (\Lambda )$ is now partially delocalized
into the leads, with the rapid fluctuations of the original spin
$\vec{S}$ now featuring as internal degrees of freedom inside the
largely static meso-spin $\vec{S}^{*} (\Lambda )$.
It is our contention that the Ward identity associated with 
spin conservation  will cause the decoherence effects in 
$\Delta
M_{\Sigma }^{(4)}$ 
to be cancelled by the vertex renormalizations in
$\Delta
M_{\Lambda }^{(4)}$,  i.e.
\[
\Delta
M_{\Sigma }^{(4)}
+\Delta
M_{\Lambda }^{(4)}= 0, 
\]
and that furthermore, these cancellations will continue order by
order. This is clearly an important point to confirm
in future work. 

We would like to end by discussing the nature of the low temperature
phase that develops in a symmetric quantum dot, where $J_{RR}=J_{LL}$.
In our previous work, we suggested that the intersite coupling terms
$J_{RL}$ are marginally irrelevant at large voltage bias, so that the
underlying low temperature physics of the large bias quantum dot is
that of a two channel Kondo model.  In such a picture, the magnetic
susceptibility would have a logarithmic divergence of the form
\begin{equation}\label{}
\chi  \propto \frac{1}{T_{K}^{*}}\ln \left(\frac{T_{K}^{*}}{T} \right)
\end{equation}
where $T_{K}^{*}\sim T_{K}^{2}/ (eV)$.   More recently, Rosch et al\cite{roschprl}
have suggested that the entry into the two channel Kondo regime
will be interrupted  at a temperature given by the
decoherence scale
\[
T_{d}\sim \frac{V}{\ln ^{2 }\left(\frac{T_{K}}{T_{d}} \right)}
\]
In the former picture, a Curie-type magnetic susceptibility would continue
all the way down to temperatures of order $T_{K}^{*}$, whereas in the
latter picture, the magnetic susceptibility would presumeably
saturate at a much higher temperature. 

The survival of a Curie susceptibility at temperatures
$T<<T_{d}$ clearly relies on the existence of a renormalized composite
``meso-spin'' $S^{*} (\Lambda )$ and the vertex cancellation effects
mentioned above. 
These same cancellation effects are intimately related
to the ability of the  meso-spin 
to transmit charge via internal fluctuations of the bare spin
$\vec{S}$, without the meso-spin undergoing fluctuations at all.
A simple physical picture of this phenomenon 
involves equal but opposite currents of electrons and holes to flowing
with the meso-spin. 
\vskip -0.1truein
\widf=0.9\columnwidth
\figa{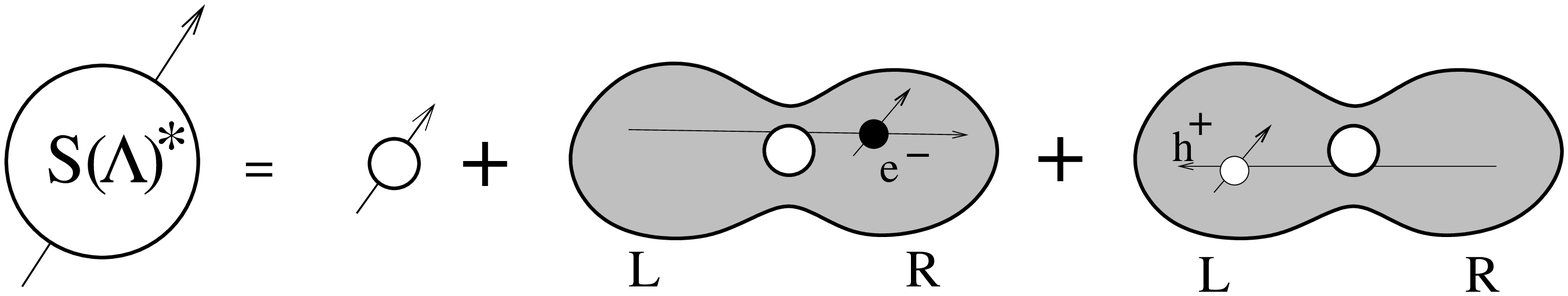}{
Physical picture of how the ``meso-spin'' 
$S^{*}(\Lambda )$ can 
carry charge from the left lead to the right lead,
through internal motion of spin within its composite structure. 
When the dot-spin migrates into the leads, a charge current
can flow without a spin current if the current of holes balances
the current of electrons.
}{fpic}\noindent 
At any given temperature, the original localized spin would then
be in a state of rapid fluctuation, but the renormalized spin would
remain essentially static until temperatures comparable with
$T_{K}^{*}$.

\subsubsection{Conclusion}\label{}

This paper has discussed the nature of the
magnetization
in the DC biased Kondo effect, highlighting it as an important central
element in the debate about whether the DC biased quantum dot
is decohered by a finite current, or whether it is able to enter
a new highly  non-equilibrium strong coupling state. 
We have proposed a slightly radical idea that the principle of virtual work
can be used in the non-equilibrium steady state by considering a
large dot-lead system  that is completely isolated, an idea that 
we showed is intimately related to our quantum reciprocity conjecture-
the idea that subclass of variables in the non-equilibrium steady
state continue  to satisfy a reciprocity relation in time.
Using these ideas, we computed an energetically defined magnetization
and showed that it has a well-defined perturbation expansion
in which the intra-lead Kondo logarithms are not cut-off
by any  finite voltage. This result supports our earlier work,
based on a point-splitting perturbation procedures.

In the final discussion, we have attempted to bring out the most
contraversial elements in the ongoing debate. 
We have argued that most of the proposed differences with our work
stem ultimately from a failure to consider the total magnetization of
the Kondo problem, rather than the local magnetization of the
dot-spin. The total magnetization is not relaxed by coupling to the
leads, and the physical susceptibility of real quantum dot always
depends on a small coupling $s$ to an external heat-bath. These difficulties
can be avoided using an energetically defined definition of the
magnetization, but they must be addressed directly if a dynamical 
calculation of the total magnetization is to be carried out. 
Finally,  we have 
proposed that so-called de-coherence effects arise because of an incomplete
and non-conserving treatment of the perturbation theory.  When
spin conservation is taken into account, cancellations take place
between the wavefunction and vertex renormalization diagrams
of the  magnetization operator that ultimately are associated with
the delocalization of the dot-spin into the leads. 
We have proposed that 
these cancellations continue into the non-equilibrium state leading
to a complete preservation of quantum coherence in the strongly biased
quantum dot.  These issues will clearly be the subject of future work.

Experimentally, it is not yet feasible to measure the magnetization of
individual quantum dots, however future improvements in the power
of scanning Hall probes may make this a feasible experiment in the future.
Another approach to the experimental examination of these phenomenon
may lie in a new configuration.  Magnetic atoms inside a 
wire between two electron baths at different voltages are exposed to 
two Fermi surface dicontinuities, and their low temperature physics
is likely to be very closely associated with the D.C. biased quantum
dot.\cite{devoret}
It may be possible by measuring the magnetic susceptibility of
pure silver wires containing rare-earth atoms with large Kondo
temperatures, such as Cerium, to actually follow the magnetic aspects
of the non-equilibrium Kondo problem.  Experimental efforts in this direction
would make an important counterparts to the theoretical debate
we have outlined above.
\\

\noindent {\bf Acknowledgments.}
\\

We  with to  particularly thank  Olivier  Parcollet and  Chris
Hooley and David Langreth 
for the many lively discussions  at the Rutgers Center for
Materials Theory that led to this paper.  
New ideas and new science can only thrive through lively debate. 
We are also grateful  to Natan Andrei,  Yshai Avishai, Premi Chandra, Leonid
Glazman,  Yehuda Goldin,  Andrew Ho,  Alexei Kaminski,  
Yigal Meir and Alexei Tsvelik for discussions related to these topics.
Finally, we are indebted to Joel Lebowitz for his suggestions and
discussions related to the concept of a non-equilibrium work function
and quantum mechanical  extensions to the Onsager  reciprocity
relation. This work was supported by DOE grant DE-FG02-00ER45790.

\section{Appendix A: Green Functions in the Larkin Ovchinikov Basis}\label{}

We begin by writing the Green function in the Larkin Ovchinikov Basis
as
\[
{\bf G}_{\alpha \beta } (1,2)= - i \langle P \Psi_{\alpha } (1)\bar
\Psi_{\beta } (2)\rangle
\]
where 
\begin{equation}\label{}
\Psi (1)\equiv (\psi _{S} (1),\psi _{A} (1))
\end{equation}
and 
\begin{equation}\label{}
\bar \Psi
(1)\equiv \Psi \dg (1)\tau _{1} = \pmatrix{\psi \dg _{A} (1)\cr
\psi \dg _{S} (1)} .
\end{equation}
Expanding this out gives
\begin{equation}\label{}
{\bf G}= -i \pmatrix{\langle P\psi _{S} (1)\psi \dg _{A}
(2)\rangle 
&
\langle P\psi _{S} (1)\psi \dg _{S} (2)\rangle \cr
\langle P\psi _{A} (1)\psi \dg _{A} (2)\rangle &
\langle P\psi _{A} (1)\psi \dg _{S} (2)\rangle }
\end{equation}
To transform this expression to its standard form, we need to expand
the path-ordered Green-functions. Let us first carry this out for the
$\langle P\psi _{S} (1)\psi \dg _{S} (2)\rangle $ term.
 Writing 
\begin{eqnarray}\label{}
\psi _{S} (1)&=&\frac{1}{\sqrt{2}}\left(\psi (1_{+})+ \psi (1_{-}) \right)\cr
\psi _{A} (1)&=&\frac{1}{\sqrt{2}}\left(\psi (1_{+})- \psi (1_{-}) \right)
\end{eqnarray}
and then expanding the path-ordering along the Keldysh contout, 
we then have
\begin{eqnarray}\label{}
-i\langle P\psi _{S} (1)\psi \dg _{S} (2)\rangle &=& 
-\frac{i }{2}\left[
\langle P \psi (1_{+})\psi \dg (2_{+})\rangle \right. \cr
+
\langle P \psi (1_{-})\psi \dg (2_{-})\rangle 
&+&\left.
\langle P \psi (1_{-})\psi \dg (2_{+})\rangle +
\langle P \psi (1_{+})\psi \dg (2_{-})\rangle 
 \right]\cr
=-\frac{i }{2}\left[
\langle T \psi (1)\psi \dg (2)\rangle \right. &+&
\langle \tilde{T} \psi (1)\psi \dg (2)\rangle 
+ \left.
\langle  [\psi (1),\psi \dg (2)]
\rangle  
 \right]\cr
&=& -i \langle [\psi (1),\ \psi \dg (2) ]
\rangle \equiv G_{K} (1,2)\nonumber
\end{eqnarray}
in a similar fashion, we see the term 
$\langle P\psi _{A} (1)\psi \dg _{A} (2)\rangle $ term vanishes:
\begin{eqnarray}\label{}
-i\langle P\psi _{A} (1)\psi \dg _{A} (2)\rangle &=& 
-\frac{i }{2}\left[
\langle P \psi (1_{+})\psi \dg (2_{+})\rangle \right. \cr
+
\langle P \psi (1_{-})\psi \dg (2_{-})\rangle 
&-&\left .
\langle P \psi (1_{-})\psi \dg (2_{+})\rangle -
\langle P \psi (1_{+})\psi \dg (2_{-})\rangle 
 \right]\cr
=-\frac{i }{2}\left[
\langle T \psi (1)\psi \dg (2)\rangle \right. &+&
\langle \tilde{T} \psi (1)\psi \dg (2)\rangle 
- \left.
\langle  [\psi (1),\psi \dg (2)]
\rangle  
 \right]\cr
&=& 0\nonumber
\end{eqnarray}
By contrast, the diagonal terms are given by
\begin{eqnarray}\label{}
-i\langle P\psi _{S} (1)\psi \dg _{A} (2)\rangle &=& 
-\frac{i }{2}\left[
\langle P \psi (1_{+})\psi \dg (2_{+})\rangle \right. \cr
\textstyle 
-
\langle P \psi (1_{-})\psi \dg (2_{-})\rangle 
\textstyle + \textstyle 
\langle P &&\hskip -0.2in\textstyle \left. \psi (1_{-})\psi \dg (2_{+})\rangle -
\langle P \psi (1_{+})\psi \dg (2_{-})\rangle 
 \right]\cr
=-\frac{i }{2}\left[
\langle T \psi (1)\psi \dg (2)\rangle \right. &-&
\left. \langle \tilde{T} \psi (1)\psi \dg (2)\rangle 
+ 
\langle  \{\psi (1),\psi \dg (2)\}
\rangle  
 \right]\cr
&=& -i\langle  \{\psi (1),\psi \dg (2)\}
\rangle  
 \theta (t_{1}-t_{2}) \cr
&\equiv& G_{R} (1,2)\nonumber
\end{eqnarray}
and
\begin{eqnarray}\label{}
-i\langle P\psi _{A} (1)\psi \dg _{S} (2)\rangle &=& 
-\frac{i }{2}\left[
\langle P \psi (1_{+})\psi \dg (2_{+})\rangle \right. \cr
\textstyle 
+
\langle P \psi (1_{-})\psi \dg (2_{-})\rangle 
\textstyle + \textstyle 
\langle P &&\hskip -0.2in\textstyle \left. \psi (1_{-})\psi \dg (2_{+})\rangle +
\langle P \psi (1_{+})\psi \dg (2_{-})\rangle 
 \right]\cr
=-\frac{i }{2}\left[
\langle T \psi (1)\psi \dg (2)\rangle \right. &-&
\langle \tilde{T} \psi (1)\psi \dg (2)\rangle 
- \left. 
\langle  \{\psi (1),\psi \dg (2)\}
\rangle  
 \right]\cr
&=&\left. +i\langle  \{\psi (1),\psi \dg (2)\}
\rangle  
 \right]\theta (t_{2}-t_{1}) \cr
&\equiv& 
G_{A} (1,2)\nonumber
\end{eqnarray}
With these manipulations, we see that we can write
\[
{\bf G} (1,2) = \pmatrix{G_{R} (1,2)&G_{K} (1,2)\cr
0 & G_{A} (1,2)
}
\]

\section{Appendix B: Feynman rules for interaction vertices}

To develop the Feynman rules for perturbation theory, we need to know
the
matrix elements of the interaction $H_{I}$ the Larkin Ovchinkov
basis.   To compute these, let us consider the simplest case of an
interaction
\begin{equation}\label{}
H_{I}= \frac{1}{2}\int d1 d2 V (1-2)\rho (1)\rho (2).
\end{equation}
where $\rho (1)= \psi \dg (1)\psi (1)$ is the density.
We can combine the density operator on the upper and lower contours
of the Keldysh path to make a symmetric and antisymmetric combination
\begin{eqnarray}\label{}
\rho _{S} (1) &=& \frac{1}{2}\left(\rho (1_{+})+\rho (1_{-}) \right)\cr
\rho _{A} (1) &=& \left(\rho (1_{+})-\rho (1_{-}) \right)\nonumber
\end{eqnarray}
The first combination is the measured value of the
particle number, whereas the second combination is the field
that couples to an external chemical potential. In the Larkin
Ovchinkov
basis, these two operators can be written 
\begin{eqnarray}\label{}
\rho _{S} (1) &=& \textstyle 
\frac{1}{2}
\left(\psi\dg  _{S} (1)
\psi  _{S} (1)+
\psi\dg  _{A} (1)
\psi  _{A} (1) 
 \right)= \frac{1}{2}\bar \Psi (1)\tau _{1}\Psi (1)\cr
\rho _{A} (1) &=& \textstyle 
\left(\psi\dg _{S} (1)
\psi  _{A} (1)+
\psi\dg  _{A} (1)
\psi  _{S} (1) 
 \right)= \bar \Psi (1)\Psi (1)\nonumber
\end{eqnarray}
so that the ``measurement vertex'',  $\tau _{1}/2$ is off-diagonal
whereas the response vertex is diagonal.
This Keldysh structure holds for any one-particle operator. 

On the Keldysh contour we can define the measurement and response
fields for $H_{I}$
as follows
\begin{eqnarray}\label{}
H^{S}_{I}&=& \frac{1}{4}\int d1 d2 V (1-2) [\rho (1_{+})\rho (2_{+})+
\rho (1_{-})\rho (2_{-})] \cr 
H^{A}_{I}&=& \frac{1}{2}\int d1 d2 V (1-2) [\rho (1_{+})\rho (2_{+})-
\rho (1_{-})\rho (2_{-})].
\end{eqnarray}
The first of these operators is used to compute the expectation value
of the interaction, but it is the 
the second operator which enters into the time-ordered exponential
and provides the scattering vertices that make up the Feynman diagrams.
These can be re-written as 
\begin{eqnarray}\label{}
H^{S}_{I}&=& \frac{1}{2}\int d1 d2 {V} (1-2) [\rho^{S} (1)\rho^{S} (2)+
\frac{1}{4}\rho ^{A}(1)\rho^{A} (2)] \cr 
H^{A}_{I}&=& \frac{1}{2}\int d1 d2 V (1-2) [\rho^{S} (1)\rho^{A} (2)+
\rho^{A} (1)\rho^{S} (2)].
\end{eqnarray}
The second term  in $H^{S}$ involves the product of two antisymmetric
operators, whose expectation value is zero. Since we only ever wish to
calculate the expectation value of $H_{I}$, we can drop these terms.
This enables us to write the interaction operators in the form
\begin{eqnarray}\label{}
H^{S}_{I}&=& \frac{1}{2}\int d1 d2 
\tilde{V}^{S}_{\lambda \lambda ',\gamma \gamma '} (1-2) 
\bar\Psi_{\lambda }
(1)\bar \Psi_{\gamma } (2)\Psi_{\gamma '}
(2)\Psi_{\lambda '} (1)
\cr 
H ^{A}_{I}&=& \frac{1}{2}\int d1 d2 \tilde{V}^{A}_{\lambda \lambda',\gamma\gamma'} (1-2) 
\bar\Psi_{\lambda }
(1)\bar \Psi_{\gamma } (2)\Psi_{\gamma '}
(2)\Psi_{\lambda '} (1)
\end{eqnarray}
where the Keldysh matrix elements for the measurement and scattering
vertices are 
\begin{eqnarray}\label{}
\tilde{V}^{S}_{\lambda \lambda',\gamma\gamma'} (1-2) &=& {
V (1-2) 
{\biggl[
\frac{\tau _{1}}{2}
\otimes \frac{\tau _{1}}{2}
\biggr]}
_{
\lambda \lambda',\gamma \gamma '
}}
\cr
\tilde{V}^{A}_{
\lambda \lambda',\gamma\gamma'} (1-2) 
&=& 
V (1-2) 
{\biggl[
\frac{\tau _{1}}{2}
\otimes \underline{1}+
\underline{1}\otimes 
\frac{\tau _{1}}{2}
\biggr]}_{\lambda \lambda',\gamma\gamma'}
\end{eqnarray}
In a Feynman diagram, each of these vertices  is associated with 
two propagators, introducing a factor $(i^{2})$. The antisymmetric
vertex also picks up a factor of $-i$ derived from the path ordered
exponential.  The final results for the scattering vertices in the
Feynman
diagrams is then 
\begin{eqnarray}\label{}
{V}^{S}_{\lambda \lambda',\gamma\gamma'} (1-2) &=& {
V (1-2) 
\biggl[
\frac{i\tau _{1}}{2}
\otimes \frac{i\tau _{1}}{2}
\biggr]_{
\lambda \lambda',\gamma\gamma'
}}\cr
{V}^{A}_{\lambda \lambda',\gamma\gamma'} (1-2) &=& V (1-2) 
\biggl[
\frac{i \tau _{1}}{2}
\otimes \underline{1}+
\underline{1}\otimes 
\frac{i\tau _{1}}{2}
\biggr]_{\lambda \lambda',\gamma\gamma'}
\end{eqnarray}
These results are quickly generalized to spin dependent interactions. 
In this paper, we are interested in the Kondo interactions of the form
\begin{eqnarray}\label{}
H_{I} = \frac{J}{2} f\dg _{\alpha }
\vec{\sigma}_{\alpha\beta } 
f_{\beta }\psi _{a }\vec{\sigma }_{ab }\psi _{b }
\end{eqnarray}
so the interaction vertex is modified by replacing
\[
V (1-2)\rightarrow  \frac{J}{2} 
\vec{\sigma}_{ab} \cdot\vec{\sigma }_{\alpha \beta }
\]
The corresponding interaction vertices are then 
\begin{eqnarray}\label{}
{V}^{S} &\equiv & 
\frac{J}{2} 
\vec{\sigma}_{ab } \cdot\vec{\sigma }_{\alpha \beta  }
\biggl[
\frac{i\tau _{1}}{2}
\otimes \frac{i\tau _{1}}{2}
\biggr]_{\lambda \lambda',\gamma\gamma'}
\cr
{V}^{A} &\equiv & \frac{J}{2} 
\vec{\sigma}_{ab } \cdot\vec{\sigma }_{\alpha \beta }
\biggl[
\frac{i \tau _{1}}{2}
\otimes \underline{1}+
\underline{1}\otimes 
\frac{i\tau _{1}}{2}
\biggr]_{\lambda \lambda',\gamma\gamma'}.
\end{eqnarray}
where for clarity, we have omitted the indices from $V^{S}$ and $V^{A}$.

\section{Appendix C: Derivation of integral identity }\label{}
In this section we prove two useful results. Firstly, that 
\bea
I_0&&=\int\frac{d\omega}{\pi}\left[f(\omega)-\frac{1}{2}\right]
\left[\frac{1}{\omega-\zeta}-\frac{1}{\omega-\zeta'}\right]
\\
=&&\frac{1}{\pi}\left[\psi\left(\frac{1}{2}+\frac{\zeta}{2\pi iT}\right)-\psi\left(\frac{1}{2}+\frac{\zeta'}{2\pi iT}\right)\right].
\eea
where $\psi(z)=d\ln \Gamma(z)/dz$ is the digamma function and second
that 
\bea
I_{1}&=&\int\frac{d\omega}{2\pi}f(\omega)\Phi(\omega)\frac{1}{\omega-\zeta}\cr
&=&\frac{1}{\pi}\left[\psi(\frac{1}{2}+\frac{\zeta}{2\pi iT})-\ln(\frac{D}{2\pi iT})\right].
\eea
where $\Phi(\omega)=D^2/(\omega^2+D^2)$ is a Lorentzian cut-off
function. 

Consider the first integral $I_{0}$. To evaluate it, we let $\zeta$ and
$\zeta'$ both lie in the upper half complex  plane. By completing 
the integral around the series of poles in the function $f(z)$ at
$z=-i\omega_n$, we obtain
\bea
I_0=&&2iT\sum_{\omega_n<0}\left(\frac{1}{i\omega_n-\zeta}-\frac{1}{i\omega_n-\zeta'}\right)\nonumber\\
=&&-\frac{1}{\pi}\sum^{\infty}_{n=1}\left(\frac{1}{n+\frac{1}{2}+\frac{\zeta}{2\pi i T}}-\frac{1}{n+\frac{1}{2}+\frac{\zeta'}{2\pi i T}}\right)
\label{I0}
\eea

But the digamma function can be expanded as a series,
\bea
\psi(z)=-C+\sum^{\infty}_{n=0}\left(\frac{1}{n+1}-\frac{1}{n+z}\right),
\eea
where $-C= \psi(1)$ is the Euler constant, so that
\bea
\psi(x)-\psi(y)=\sum_{n=0}^{\infty}\left(\frac{1}{n+y}-\frac{1}{n+x}\right).
\eea
Substituting this into (\ref{I0} ), we then obtain
\begin{equation}\label{I0X}
I_{0}
=\frac{1}{\pi}\left[\psi\left(\frac{1}{2}+\frac{\zeta}{2\pi iT}\right)-\psi\left(\frac{1}{2}+\frac{\zeta'}{2\pi iT}\right)\right]
\end{equation}
Let us turn to the second integral
\bea
I_{1}=\int\frac{d\omega}{\pi}\left[ f(\omega)-\frac{1}{2}\right]\Phi(\omega)\frac{1}{\omega-\zeta},
\eea
where $\Phi(\omega)=D^2/(\omega^2+D^2)$. Expanding out the cut-off function, writing
\bea
\Phi(\omega)=\frac{D}{2i}\left[\frac{1}{\omega-iD}-\frac{1}{\omega+iD}\right],
\nonumber
\eea
then
\bea
I_{1}=\frac{D}{2i}\int\frac{d\omega}{\pi}\left[ f(\omega)-\frac{1}{2}\right]\left[\frac{1}{\omega-\zeta}\frac{1}{\omega-iD}+\{D\rightarrow -D\}\right]
\nonumber
\eea
in the limit where $D\gg|\zeta|$, we have
\bea
I_{1}=\int\frac{d\omega}{2\pi}\left[f(\omega)-\frac{1}{2}\right]\left[(\frac{1}{\omega-\zeta}-\frac{1}{\omega-iD})+\{D\rightarrow -D\}\right].
\nonumber
\eea
But the second term is equal to the first one, because
\bea
\int \frac{d\omega}{2\pi}\left[f(\omega)-\frac{1}{2}\right]\frac{1}{\omega-iD}=\int\frac{d\omega}{2\pi}\left[f(\omega)-\frac{1}{2}\right]\frac{1}{\omega+iD}
\nonumber
\eea
(the difference between these two integrals is the integral of an odd, with an even function, so it vanishes), so that
\bea
I_{1}=\int\frac{d\omega}{\pi}\left[f(\omega)-\frac{1}{2}\right]\left[\left(\frac{1}{\omega-\zeta}-\frac{1}{\omega-iD}\right)\right].
\eea
Using Eq.(\ref{I0X}), we get
\bea
&&\int\frac{d\omega}{\pi}\left[f(\omega)-\frac{1}{2}\right]\Phi(\omega)\frac{1}{\omega-\zeta}\nonumber\\
=&&\frac{1}{\pi}\left[\psi(\frac{1}{2}+\frac{\zeta}{2\pi i T})-\psi(\frac{1}{2}+\frac{D}{2\pi T})\right]\nonumber\\
=&&\frac{1}{\pi}\left[\psi(\frac{1}{2}+\frac{\zeta}{2\pi i T})-\ln(\frac{D}{2\pi T})\right].
\eea
Finally, since
\bea
\int\frac{d\omega}{2\pi}\Phi(\omega)\frac{1}{\omega-\zeta}=\frac{i}{2},
\eea
we reach
\bea
\int\frac{d\omega}{2\pi}f(\omega)\Phi(\omega)\frac{1}{\omega-\zeta}=\frac{1}{\pi}\left[\psi(\frac{1}{2}+\frac{\zeta}{2\pi iT})-\ln(\frac{D}{2\pi iT})\right].
\eea

\section{Appendix D: Derivation of charge current and conductance }\label{}
In this section, we derive an 
expression for charge current to the second order of $J$ using the
interaction energy $\langle H_{I}\rangle $. 
While the Onsager reciprocity relation does not hold for charge
current and interaction energy, we are able to relate the current 
to the derivative of the interaction energy with respect to the
interaction energy\cite{colemanmao:01}, 
\begin{eqnarray}
I &=& \left.-\frac{\partial}{\partial A}\right|_{A=0}
\langle H_{I}\rangle .
\end{eqnarray}
where the above quantities are to be calculated by including
a vector potential $A$ into the scattering potential (but not into the
measurement vertex of $H_{I}$), by replacing
\[
J_{RL}\rarrow J_{RL}e^{{-i\frac{eA}{\hbar}}}, \qquad 
J_{LR}\rarrow J_{LR}e^{{i\frac{eA}{\hbar}}}.
\]
After making this 
gauge transform we obtain the following expression for the interaction
energy to second order  (
at zero magnetic field $B=0$),
\begin{eqnarray}
\langle H^{(2)}_I\rangle =
&& 3J_{LR}^2\int\frac{d\epsilon}{2\pi}\nonumber\\
&&\left\{
\left[
\pi^R_{L}(\epsilon)\pi^K_{R}(\epsilon)+\pi^K_{L}(\epsilon)\pi^A_{R}(\epsilon)
\right]
e^{-\frac{ieA}{\hbar}}
\right.
\cr
&&
+\left.
\left[
\pi^R_{R}(\epsilon)\pi^K_{L}(\epsilon)+\pi^K_{R}(\epsilon)\pi^A_{L}(\epsilon)
\right]
e^{\frac{ieA}{\hbar}}
\right\}
\end{eqnarray}
where the factor 3 comes from summation of spins, Eq.(\ref{spins}).
By differentiating with respect to the vector potential, we obtain
\begin{eqnarray}
I^{(2)}=&&3J_{LR}^2
\frac{i e}{\hbar}
\int
\frac{d\epsilon}{2\pi}
\left\{
-\left[\pi^K_{L}(\epsilon) \left( \pi^R_{R}(\epsilon)-\pi^A_{R}(\epsilon)\right)\right]\right.\nonumber\\&&
\left.
+\left[\pi^K_{R}(\epsilon) 
\left( \pi^R_{L}(\epsilon)-\pi^A_L(\epsilon)
\right)
\right] 
\right\}
\end{eqnarray}
By using Eq.(\ref{hard1}) and (\ref{hard2}), 
the zero temperature current is then 
\bea
I^{(2)}&&=3J_{LR}^2\frac{e\pi \rho
^{2}}{\hbar}\int\frac{d\epsilon}{2\pi}
\rm{Im}\left[\psi(\frac{1}{2}+\frac{\epsilon+\mu_R}{2\pi i
T})-\psi(\frac{1}{2}+\frac{\epsilon+\mu_L}{2\pi i T})\right]
\cr
&&={3J_{LR}^2\frac{e^2}{\hbar}\pi\rho^2} V,
\eea
hence
\bea
G=\frac{{3e^2\pi}}{\hbar}J_{LR}^2{\rho^2}.
\eea

\end{document}